\documentclass[12pt,a4paper,notitlepage]{article}

\usepackage{amsfonts}
\usepackage{amsmath}
\usepackage{amssymb}
\usepackage{amsthm}
\usepackage{mathtools}
\usepackage{bm}
\usepackage{color}
\usepackage[T1]{fontenc}
\usepackage{graphicx}
\usepackage{booktabs}
\usepackage{hyperref}
\usepackage{mathrsfs}
\usepackage{multirow}
\usepackage[bibstyle=ieee,citestyle=numeric-comp,sorting=none]{biblatex} 
\usepackage[format=hang]{subcaption}
\usepackage{times}
\usepackage{upgreek}
\usepackage{titlesec}
\usepackage{wrapfig}
\usepackage[font=footnotesize,labelfont=bf]{caption}
\usepackage{booktabs}
\usepackage{tabularx}
\usepackage{array}
\newcolumntype{P}[1]{>{\centering\arraybackslash}p{#1}}
\usepackage{enumitem}
\setlist[itemize]{noitemsep, nosep}
\usepackage{physics}
\usepackage[title,toc,titletoc]{appendix}
\usepackage{hyphenat}
\usepackage{xfrac}  

\bibliography{literature}

\renewcommand{\footnoterule}{
  \hrule width \textwidth
  \kern 6pt
}

\DeclareMathOperator{\sgn}{sgn}

\titlelabel{\thetitle.\;}

\date{}

\graphicspath{{./figures/}}

\begin{document}


\author{\vspace{-0.5cm}
\parbox{\linewidth}{\centering
\large {Hooman Danesh${}^{{\,a},}$\footnote{Corresponding author: hooman.danesh@rwth-aachen.de} \ , Lisamarie Heußen${}^{\,a}$, Francisco J. Montáns${}^{\,b,c}$, Stefanie Reese${}^{\,a}$, Tim Brepols${}^{\,a}$}\\[0.5cm]
\footnotesize{\em ${}^{a}$ Institute of Applied Mechanics, RWTH Aachen University, Mies-van-der-Rohe-Str. 1, 52074 Aachen, Germany}\\
\footnotesize{\em ${}^{b}$ ETS de Ingeniería Aeronáutica y del Espacio, Universidad Politécnica de Madrid, Pza Cardenal Cisneros 3, 28040 Madrid, Spain}\\
\footnotesize{\em ${}^{c}$
Department of Mechanical and Aerospace Engineering, Herbert Wertheim College of Engineering, University of Florida, Gainesville, FL 32611, USA}
}}

\title{\vspace{-1.0cm} \LARGE A two-scale computational homogenization approach for elastoplastic truss-based lattice structures}

\maketitle 

\hrulefill

\small
{\bf Abstract}\\
The revolutionary advancements in metal additive manufacturing have enabled the production of alloy-based lattice structures with complex geometrical features and high resolutions. This has encouraged the development of nonlinear material models, including plasticity, damage, etc., for such materials. However, the prohibitive computational cost arising from the high number of degrees of freedom for engineering structures composed of lattice structures highlights the necessity of homogenization techniques, such as the two-scale computational homogenization method. In the present work, a two-scale homogenization approach with on-the-fly exchange of information is adopted to study the elastoplastic behavior of truss-based lattice structures. The macroscopic homogenized structure is represented by a two-dimensional continuum, while the underlying microscale lattices are modeled as a network of one-dimensional truss elements. This helps to significantly reduce the associated computational cost by reducing the microscopic degrees of freedom. The microscale trusses are assumed to exhibit an elastoplastic material behavior characterized by a combination of nonlinear exponential isotropic hardening and linear kinematic hardening. Through multiple numerical examples, the performance of the adopted homogenization approach is examined by comparing forces and displacements with direct numerical simulations of discrete structures for three types of stretching-dominated lattice topologies, including triangular, X-braced and X-Plus-braced unit cells. Furthermore, the principle of scale separation, which emphasizes the need for an adequate separation between the macroscopic and microscopic characteristic lengths, is investigated. It is demonstrated that by employing a sufficient number of lattice structures, the homogenization framework results in highly precise solutions during loading, unloading and reverse loading scenarios, exhibiting a high degree of agreement with full-scale simulations in both the elastic and elastoplastic regions.


\newpage
\normalsize

\section{Introduction}
\label{sec:1}

With the development of additive manufacturing technologies, the production of structures with complex geometrical features has been considerably simplified. This allows for the design and fabrication of materials with exceptional properties, such as appealing stiffness-to-weight ratios \cite{zheng2014ultralight}, negative Poisson's ratio \cite{babaee20133d} and superb thermal \cite{you2021design} or acoustic \cite{gazzola2021mechanics} insulation performance (see, e.g., \cite{kadic20193d,fischer2020mechanical,fan2021review} for review). Many of these 3D-printed structures fall under the category of metamaterials, a term coined by Walser \cite{walser2000metamaterials} to describe materials with capabilities exceeding those of conventional ones.

Among different classes of metamaterials (e.g., mechanical, acoustic, photonic, thermal, etc.), truss-based mechanical metamaterials are the ones that are formed from a distribution of thin struts to achieve mechanical qualities of interest. In the case of regular truss networks, truss-based metamaterials are generated from a periodic arrangement of a single lattice structure (i.e., unit cell) \cite{fleck2010micro,meza2017reexamining,benedetti2021architected}. Based on the topology of this unit cell, such as triangular, hexagonal and kagome topologies in two dimensions \cite{fleck2010micro} or octet-truss, cuboctahedron, BCC and FCC lattice families in three dimensions \cite{benedetti2021architected}, a variety of tailored mechanical properties becomes possible.

When it comes to the computational modeling of metamaterials, direct numerical simulations (DNS) of real-world engineering structures made of these discrete structures are highly computationally demanding. Therefore, the development of homogenized continuum models to extract the effective response of such structures appears to be absolutely inevitable. Accordingly, a number of miscellaneous homogenization approaches, ranging from analytical methods to numerical and hybrid ones, have been introduced (e.g., see \cite{kochmann2019multiscale}). Analytical methods, which are based on the pre-computation of the effective response of the lattice structure, lose their potential as far as complex topologies or nonlinear behaviors are concerned. As a key remedy, computational homogenization techniques \cite{smit1998prediction,feyel2000fe2,kouznetsova2001approach} can be employed to obtain the homogeneous response of larger scale (i.e., macroscale) structures by concurrently solving the smaller scale (i.e., microscale) unit cell boundary value problems (BVPs). Based on the solution technique, e.g., finite element (FE) or fast Fourier transform (FFT), employed at each of these two adequately separated scales, the two-scale computational homogenization method is referred to under different names, such as $\mathrm{FE^2}$ \cite{feyel2000fe2,feyel1999multiscale,feyel2003multilevel} or FE-FFT \cite{spahn2014multiscale, kochmann2016two,gierden2022review}. Although the FE-FFT approach has been shown to be competitively efficient for two-scale homogenization of composites \cite{spahn2014multiscale} or polycrystals \cite{gierden2021efficient}, the applicability of such FFT-based homogenization approaches for lattice structures is challenging due to an existing infinite stiffness contrast \cite{danesh2023challenges}. Therefore, $\mathrm{FE^2}$-type methods seem to be a more straightforward two-scale homogenization approach for lattice-based materials \cite{vigliotti2012linear,nguyen2014computational,wu2023second}.

When dealing with the $\mathrm{FE^2}$ approach, the computational cost is significantly influenced by the discretization of both the macroscale structure and the microscale lattice. This is because a representative volume element (RVE) needs to be solved at each macroscopic integration point to numerically evaluate the constitutive relation. While it is true that continuum solid elements offer more realistic results, it is possible to represent lattice structured materials by structural elements, such as truss or beam elements \cite{gibson1982mechanics,deshpande2001effective,deshpande2001foam,wadley2003fabrication}. As a direct consequence, when such structural elements are employed within the scope of two-scale homogenization \cite{vigliotti2012linear,desmoulins2017local,glaesener2019continuum}, the degrees of freedom (DoFs) for the fine-scale unit cell will considerably reduce from a magnitude of hundreds or thousands to a mere tens, resulting in a significant reduction of the computational effort. Moreover, thanks to the existence of analytical consistent tangents, the computational cost will further decrease in comparison to the $\mathrm{FE^2}$ approaches with continuum solid elements \cite{feyel2000fe2,feyel1999multiscale}, which require numerical derivatives \cite{miehe1996numerical}.

Vigliotti et al. \cite{vigliotti2012linear} took advantage of beam elements within a two-scale  $\mathrm{FE^2}$-type approach to determine the elastic macroscopic properties of stretching- and bending-dominated lattice structures (i.e., pin-jointed and rigid-jointed lattices, respectively) within a geometrically linear framework. In order to consider size and localization effects, Weeger \cite{weeger2021numerical} employed a second-gradient linear elastic theory to represent the macroscopic homogenized continuum, while the microscale lattices were modeled as shear deformable-beam elements. Nonetheless, as recent additive manufacturing advances have made it possible to fabricate lattice-based materials with a diverse range of base constituents, a compelling need to develop nonlinear models for such materials has arisen in recent years. There exists a number of works focusing on two-scale homogenization frameworks for truss-based lattice structures, considering geometrical and/or material nonlinearities. Nonlinear buckling of lattice materials subjected to large strains has been investigated in \cite{vigliotti2014non} using a computational homogenization approach. Examining the influence of the RVE size revealed that it only affects post-bifurcation behavior in loading paths, emphasizing the necessity of preliminary investigations for accurate RVE selection. A second-gradient two-scale finite deformation model, characterized by a generalized macroscale continuum with an intrinsic length scale, was developed for two-dimensional \cite{glaesener2019continuum} and three-dimensional \cite{glaesener2020continuum} linear elastic beam networks. In this approach, in addition to translational DoFs, rotational DoFs of microscale beams were also coupled with the macroscale, leading to a macroscale rotation field similar to the micropolar theory \cite{eringen1966linear}. It was shown that using a second-gradient homogenization method enhances local accuracy in cases of localization compared to a first-gradient scheme, but this improvement does not notably affect the overall global response. Subsequently, the same approach \cite{glaesener2019continuum,glaesener2020continuum} was extended to beam networks with material nonlinearity, including viscoelastic behavior \cite{glaesener2021viscoelastic}.

Yet, due to the recent advancements in metal 3D printing, a substantial need for developing homogenization frameworks capable of accounting for nonlinear plastic behavior arises. Vigliotti et al. \cite{vigliotti2012stiffness} extended their model to three dimensions and extracted the von Mises surfaces for plastic yielding of different open and closed cell lattices. However, their study was restricted to only achieving the proportional limits of lattices, and the post-yielding deformations, characterized by plastic hardening, were not considered. By virtue of its effective performance in terms of both accuracy and efficiency, the aforementioned homogenization framework \cite{vigliotti2012linear,vigliotti2012stiffness,vigliotti2014non} renders it both persuasive and methodologically sound to be extended for elastoplastic lattice structures.

In the present work, following the same approach as in \cite{vigliotti2012linear,vigliotti2012stiffness,vigliotti2014non}, an on-the-fly two-scale computational homogenization approach for elastoplastic truss-based metamaterials is developed. Compared to \cite{vigliotti2012linear,vigliotti2012stiffness}, where the effective macroscopic properties were extracted by solely analyzing a single unit cell without concurrently solving the macroscopic and microscopic BVPs, a completely coupled two-scale solution procedure, similar to \cite{vigliotti2014non}, is performed in the present work due to the existing material nonlinearity. The macroscopic homogenized structure is expressed by a two-dimensional continuum, which can be discretized by any element of choice, such as the four-node bi-linear quadrilateral elements considered in this study. The microscale struts are modeled using geometrically linear two-node structural truss elements, which are perfectly suitable for stretching-dominated (i.e., pin-jointed) lattices. An elastoplastic material behavior with combined nonlinear exponential isotropic hardening and linear kinematic hardening is considered for lattice struts. In comparison with \cite{vigliotti2012stiffness}, where the main focus was the buckling and yield strength of lattice structures, the nonlinear post-yielding response is  accounted for in the present study as well. By comparing the homogenization results with full-resolution simulations, the employed approach is shown to be very suitable for a variety of stretching-dominated lattice structures, including triangular, X-braced and X-Plus-braced (hereinafter referred to as XP-braced) lattices. The principle of scale separation \cite{geers2010multi} is examined, revealing that an adequate number of lattice structures are required in order for the homogenization scheme to provide accurate enough solutions.

The paper is organized as follows. Section \ref{sec:2} is concerned with the theoretical foundation of the two-scale homogenization model, including macroscale and microscale BVPs as well as the bridging of these two scales. Section \ref{sec:3} briefly addresses the FE formulation, followed by the constitutive model in Section \ref{sec:4}, including the elastoplastic behavior of the microscale lattice structure, characterized by a combined nonlinear isotropic and linear kinematic hardening rule. The algorithmic implementation of the two-scale scheme is dealt with in Section \ref{sec:5}. Section \ref{sec:6} focuses on several numerical examples, where comparisons of the results obtained from the homogenized model and the full-scale discrete simulation are performed for different pin-jointed lattice topologies. Finally, concluding remarks and future ideas are given in Section \ref{sec:7}.
\section{Two-scale homogenization}
\label{sec:2}

\begin{figure}[h]
    \centering
    \includegraphics[keepaspectratio=true, width=0.5\linewidth]{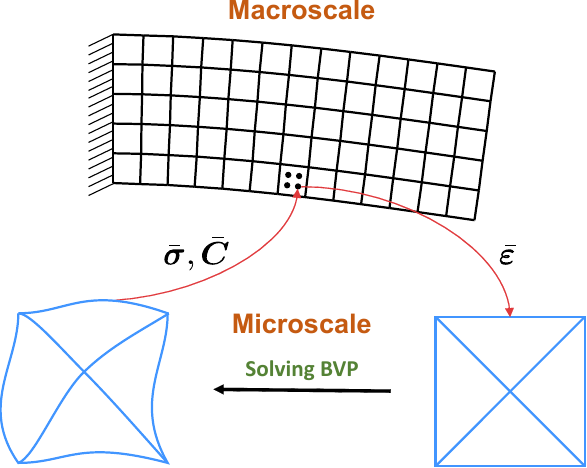}
    \caption{Schematic representation of the two-scale computational homogenization approach. At each integration point of the macroscale BVP, the macroscopic strain field $\bar{\bm{\varepsilon}}$ is transferred to the microscale, where it is used to solve another BVP for a unit cell composed of truss elements. The homogenized stress $\bar{\bm{\sigma}}$ and tangent $\bar{\bm{C}}$ are then transferred back to the macroscale.}
    \label{fig:1}
\end{figure}

Based on the two-scale computational homogenization scheme \cite{kouznetsova2001approach,geers2010multi}, two BVPs need to be defined at two adequately separated scales, widely named the macroscale and the microscale, even if the smaller scale does not possess micrometer dimensions. Here, the macroscale BVP follows a conventional solid mechanics problem except for the fact that there exists no explicit stress-strain relation, and such a material behavior is obtained by solving a microscale BVP. As illustrated in Figure \ref{fig:1}, at each integration point of the macroscale body, the macroscopic strain field $\bar{\bm{\varepsilon}}$ is transferred to the microscale, where it is employed to apply the microscopic boundary conditions (BCs). Subsequently, the homogenized stress field $\bar{\bm{\sigma}}$ and tangent stiffness $\bar{\bm{C}}$ are computed by exploiting the volume average of the microscale quantities. The succeeding sections are devoted to the description of the macroscale and microscale BVPs and the transition of information between these two scales.

\subsection{Macroscale}
\label{subsec:2.1}

The macroscale continuum is governed by the general equations of small-strain solid mechanics, including the balance of linear momentum in the absence of body forces,

\begin{equation}
\label{eq:1}
    \div \bar{\bm{\sigma}} = \bm{0},
\end{equation}

and the kinematic equations,

\begin{equation}
\label{eq:2}
    \bar{\bm{\varepsilon}} = \grad^{s} \bar{\bm{u}},
\end{equation}

where $\bar{\bm{\sigma}}$ and $\bar{\bm{\varepsilon}}$ are the macroscopic Cauchy stress and small strain tensors, respectively, and $\bar{\bm{u}}$ denotes the displacement vector. $\bm{\nabla}$ is the differential operator vector, and $\bm{\nabla}^{s}$ denotes its symmetric part. It is also to be noted that the macroscopic variables are denoted by $\bar{\left( \cdot \right)}$, implying that they represent the averaged homogenized quantities. Eventually, the local variation of deformation work at the macroscale $\delta \bar{W}$, which will be used for the scale bridging (Section \ref{subsec:2.3}), can be expressed as

\begin{equation}
\label{eq:3}
    \delta \bar{W} =  \bar{\bm{\sigma}} : \delta \bar{\bm{\varepsilon}}.
\end{equation}

In the following, matrix notation will be used, where for the 2D case, the strain and stress tensors are presented in the form of $\bar{\bm{\varepsilon}} \in \mathbb{R}^{2\times2}$ and $\bar{\bm{\sigma}} \in \mathbb{R}^{2\times2}$ matrices, respectively. For subsequent use, it is more convenient to take advantage of the underlying symmetry in the stress and strain tensors and present them as arrays based on Voigt notation, denoted by $\hat{\bar{\bm{\varepsilon}}} \in \mathbb{R}^{3}$ and $\hat{\bar{\bm{\sigma}}} \in \mathbb{R}^{3}$, as

\begin{equation}
\label{eq:4}
    \bar{\bm{\varepsilon}} = \begin{pmatrix}
    \bar{\varepsilon}_{xx} & \bar{\varepsilon}_{xy}\\
    \bar{\varepsilon}_{xy} & \bar{\varepsilon}_{yy}
    \end{pmatrix},
    \quad
    \hat{\bar{\bm{\varepsilon}}} = \begin{pmatrix}
    \bar{\varepsilon}_{xx}\\
    \bar{\varepsilon}_{yy}\\
    \bar{\gamma}_{xy} 
    \end{pmatrix},
    \quad
    \bar{\bm{\sigma}} = \begin{pmatrix}
    \bar{\sigma}_{xx} & \bar{\sigma}_{xy}\\
    \bar{\sigma}_{xy} & \bar{\sigma}_{yy}
    \end{pmatrix},
    \quad
    \hat{\bar{\bm{\sigma}}} = \begin{pmatrix}
    \bar{\sigma}_{xx}\\
    \bar{\sigma}_{yy}\\
    \bar{\sigma}_{xy} 
    \end{pmatrix},
\end{equation}

where $\bar{\gamma}_{xy} = 2\bar{\varepsilon}_{xy}$. This helps to write Eq. \ref{eq:3} as a scalar product in the form

\begin{equation}
\label{eq:5}
    \delta \bar{W} =  \hat{\bar{\bm{\sigma}}}^T \delta  \hat{\bar{\bm{\varepsilon}}}.
\end{equation}

\subsection{Microscale}
\label{subsec:2.2}

\begin{figure}[ht]
    \centering
    \includegraphics[keepaspectratio=true, width=\linewidth]{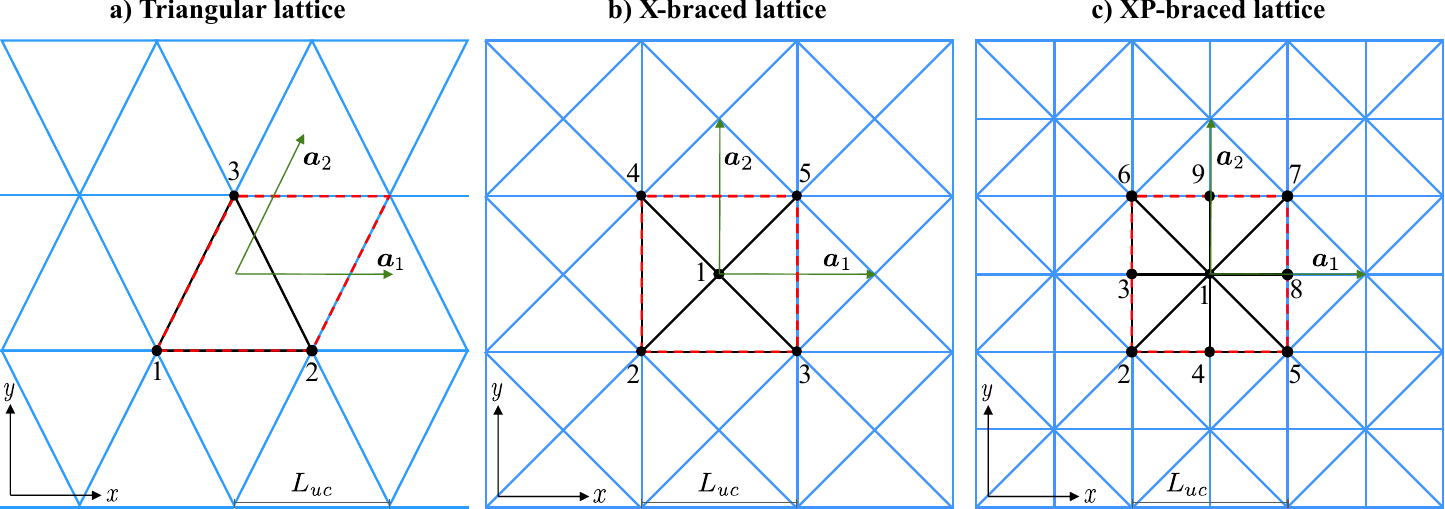}
    \caption{Unit cell topology and node numbering for three types of stretching-dominated lattice structures: \textbf{a)} triangular lattice, \textbf{b)} X-braced lattice and \textbf{c)} XP-braced lattice. Dashed lines show the the boundaries of the unit cell, and the translational periodic vectors are denoted by $\bm{a}_{1}$ and $\bm{a}_{2}$.}
    \label{fig:2}
\end{figure}

In order to employ structural truss elements for the representation of microscale lattice structures in the two-scale homogenization scheme, the approach presented in \cite{vigliotti2012linear,vigliotti2012stiffness,vigliotti2014non} is followed. This method is briefly revisited here for the special case of triangular lattice structure (Figure \ref{fig:2}a), and the same derivation for the X-braced (Figure \ref{fig:2}b) and XP-braced (Figure \ref{fig:2}c) unit cells can be followed in Appendix \ref{app:a}. Considering a single unit cell, the nodes of the cell can be classified into two categories, namely the \textit{internal nodes} and the \textit{boundary nodes}. Internal nodes exist at intersections of the elements of the present unit cell, and boundary nodes connect the nodes of a given unit cell to neighbouring cells. The periodicity of the unit cells necessitates the alignment of the boundary nodes along the periodic vectors and the existence of a minimum of one matching node on the opposing boundary. Moreover, the cell nodes can be categorized into the two groups of \textit{independent nodes} and \textit{dependent nodes}, which enables computing the positions of the dependent nodes from the positions of the independent nodes and the periodic vectors. It is to be noted that all the internal nodes fall into the category of independent nodes since it is impossible to compute the position of any other node by translating them along the periodic vectors or their combination. In general, the position vector of the $i^{\mathrm{th}}$ dependent node, denoted by $\bm{r}_i$, can be written in terms of the position vector of the $j^{\mathrm{th}}$ independent node, labelled by $\bm{r}_j$, and the translational periodic vectors $\bm{a}_k$ as

\begin{equation}
\label{eq:6}
    \bm{r}_{i} = \bm{r}_{j} + \sum_{k=1}^{N_{dim}} m_{k}\bm{a}_{k}.
\end{equation}

In the above equation, $j \in \left\{ 1, \hdots , N_{ind} \right\}$ with $N_{ind}$ being the number of independent nodes and $i \in \left\{ N_{ind} + 1, \hdots , N_{tot} \right\}$ with $N_{tot}$ defining the total number of nodes. It is to be noted that the number of dependent nodes $N_{dep}$ is obtained as $N_{dep} = N_{tot} - N_{ind}$. Moreover, $m_{k} \in \left\{ -1, 0, 1 \right\}$, where $k \in \left\{ 1, \hdots , N_{dim} \right\}$ with $N_{dim}$ denoting the spatial dimension (i.e., 2 or 3). The total number of DoFs and the number of independent DoFs for a unit cell are defined as $n_{tot} = N_{dim} N_{tot}$ and $n_{ind} = N_{dim} N_{ind}$, respectively.

Considering the triangular lattice, as illustrated in Figure \ref{fig:2}a, the lattice is composed of one independent node (i.e., node 1) and two dependent nodes (i.e., nodes 2 and 3). Then, the position of the dependent nodes 2 and 3 can be written in terms of the position of the independent node 1 and the periodic vectors $\bm{a}_{1}$ and $\bm{a}_{2}$. These periodic vectors show the directions in which the unit cell is periodically distributed. The nodal positions of dependent nodes for the triangular lattice are written as

\begin{align}
\begin{split}
\label{eq:7}
    & \bm{r}_{2} = \bm{r}_{1} + \bm{a}_{1},\\
    & \bm{r}_{3} = \bm{r}_{1} + \bm{a}_{2}.
\end{split}
\end{align}

The periodic vectors $\bm{a}_{k}$ will deform as a result of the imposed uniform macroscopic strain field $\bar{\bm{\varepsilon}}$ \cite{asaro2006mechanics} as

\begin{equation}
\label{eq:8}
    \bm{a}'_{k} = \left( \bm{\mathrm{I}} + \bar{\bm{\varepsilon}} \right) \bm{a}_{k},
\end{equation}

where $\bm{a}'_{k}$ denote the deformed periodic vectors. Subsequently, the position of the $i^{\mathrm{th}}$ dependent node of the unit cell after deformation is obtained as follows:

\begin{equation}
\label{eq:9}
    \bm{r}'_{i} = \bm{r}'_{j} + \sum_{k=1}^{N_{dim}} m_k \bm{a}'_k = \bm{r}'_{j} + \sum_{k=1}^{N_{dim}} m_k \left( \bm{\mathrm{I}} + \bar{\bm{\varepsilon}} \right) \bm{a}_{k}
\end{equation}

Eventually, by subtracting the reference position of the $i^{\mathrm{th}}$ dependent node from its current position (i.e., subtracting Eq. \ref{eq:6} from Eq. \ref{eq:9}), the displacement vector of the given node $\bm{d}_{i}$ is obtained as

\begin{equation}
\label{eq:10}
    \bm{d}_{i} = \bm{r}'_i - \bm{r}_i = \bm{d}_{j} + \sum_{k=1}^{N_{dim}} m_{k} \bar{\bm{\varepsilon}} \bm{a}_{k}. 
\end{equation}

Regarding the triangular lattice, the nodal displacements will have the form

\begin{align}
\begin{split}
\label{eq:11}
    & \bm{d}_{2} = \bm{d}_{1} + \bar{\bm{\varepsilon}} \bm{a}_{1},\\
    & \bm{d}_{3} = \bm{d}_{1} + \bar{\bm{\varepsilon}} \bm{a}_{2}.
\end{split}
\end{align}

For the sake of compact representation of the equations and comprehensible programming flow, the aforementioned equations are written in an array form. To do so, the DoFs corresponding to all the nodes of the unit cell, including both the independent and dependent ones, are gathered in the array $\bm{d} \in \mathbb{R}^{n_{tot}}$, and the DoFs of all the independent nodes are collected in the array $\bm{d}_0 \in \mathbb{R}^{n_{ind}}$. Taking Eq. \ref{eq:10} into consideration, the array of nodal DoFs $\bm{d}$ can be obtained in terms of the array of independent DoFs $\bm{d}_0$ and the array of macroscopic strain components $\hat{\bar{\bm{\varepsilon}}}$ as

\begin{equation}
\label{eq:12}
    \bm{d} = \bm{B}_{0} \bm{d}_{0} + \bm{B}_{e} \hat{\bar{\bm{\varepsilon}}}.
\end{equation}

In Eq. \ref{eq:12}, $\bm{B}_{0} \in \mathbb{R}^{n_{tot} \times n_{ind}}$ and $\bm{B}_{e} \in \mathbb{R}^{n_{tot} \times 3\left(N_{dim}-1\right)}$ are block matrices attained from the topology of the unit cell, where $\bm{B}_{0}$ is an array composed of zero and identity matrices, and $\bm{B}_{e}$ is the array that maps the macroscopic strain components to the nodal displacements. With regard to the triangular lattice, the arrays $\bm{d} \in \mathbb{R}^{6}$, $\bm{d}_0 \in \mathbb{R}^{2}$, $\bm{B}_{0} \in \mathbb{R}^{6\times2}$ and $\bm{B}_{e} \in \mathbb{R}^{6\times3}$ are defined as follows:

\begin{equation}
\label{eq:13}
    \bm{d} = \begin{pmatrix}
    \bm{d}_{1}\\
    \bm{d}_{2}\\
    \bm{d}_{3}
    \end{pmatrix},
    \quad
    \bm{d}_{0} = \begin{pmatrix}
    \bm{d}_{1}
    \end{pmatrix},
    \quad
    \bm{B}_{0} = \begin{pmatrix}
    \bm{I}\\
    \bm{I}\\
    \bm{I}
    \end{pmatrix},
    \quad
    \bm{B}_{e} = \begin{pmatrix}
    \bm{0}\\
    \bm{B}_{{e}_2}\\
    \bm{B}_{{e}_3}
    \end{pmatrix}
\end{equation}

Expanding Eqs. \ref{eq:11} for the $x$ and $y$ displacements as

\begin{align}
\begin{split}
\label{eq:14}
    & d_{2x} = d_{1x} + \bar{\varepsilon}_{xx} a_{1x} + \dfrac{1}{2} \bar{\gamma}_{xy} a_{1y},
      \quad
      d_{2y} = d_{1y} + \dfrac{1}{2} \bar{\gamma}_{xy} a_{1x} + \bar{\varepsilon}_{yy} a_{1y},\\[1em]
    & d_{3x} = d_{1x} + \bar{\varepsilon}_{xx} a_{2x} + \dfrac{1}{2} \bar{\gamma}_{xy} a_{2y},
      \quad
      d_{3y} = d_{1y} + \dfrac{1}{2} \bar{\gamma}_{xy} a_{2x} + \bar{\varepsilon}_{yy} a_{2y},
\end{split}
\end{align}

the sub-matrices of $\bm{B}_{e}$ (i.e., $\bm{B}_{{e}_2} \in \mathbb{R}^{2\times3}$ and $\bm{B}_{{e}_3} \in \mathbb{R}^{2\times3}$) can be extracted as

\begin{equation}
\label{eq:15}
    \bm{B}_{e_{2}} = \begin{pmatrix}
    a_{1x} & 0 & \frac{1}{2} a_{1y} \\
    0 & a_{1y} & \frac{1}{2} a_{1x}
    \end{pmatrix},
    \quad
    \bm{B}_{e_{3}} = \begin{pmatrix}
    a_{2x} & 0 & \frac{1}{2} a_{2y} \\
    0 & a_{2y} & \frac{1}{2} a_{2x}
    \end{pmatrix},
\end{equation}

where $a_{kx}$ and $a_{ky}$ are the components of the periodic vectors $\bm{a}_{k}$ and can be written in terms of the side length of the unit cell $L_{uc}$. In the present work, it is considered that the triangular lattice is composed of an isosceles triangle with the base (i.e., the horizontal side in Figure \ref{fig:2}a) and the height being the same length and equal to $L_{uc}$. Therefore, the periodic vectors for the triangular lattice can be expressed as

\begin{equation}
\label{eq:16}
    \bm{a}_{1} = L_{uc} \begin{pmatrix}
    1\\
    0
    \end{pmatrix},
    \quad
    \bm{a}_{2} = L_{uc} \begin{pmatrix}
    \sfrac{1}{2}\\
    1
    \end{pmatrix}.
\end{equation}

As shown through the aforementioned equations, the independent DoFs $\bm{d}_{0}$ are the primary unknowns of the system, which can be solved for by means of the periodic equilibrium for the nodal forces of the unconstrained unit cell. More precisely, assuming that the periodicity of lattices is held during deformation, the equilibrium of the unit cell under the influence of its neighboring cells can be expressed in terms of the nodal forces of the given unit cell itself (see \cite{vigliotti2012linear} for more details). For instance, in the case of a triangular lattice, equilibrium requires that

\begin{equation}
\label{eq:17}
    \bm{f}_{1} + \bm{f}_{2} + \bm{f}_{3} = \bm{0}.
\end{equation}

Gathering all the nodal forces of the unit cell in the array $\bm{f} \in \mathbb{R}^{n_{tot}}$, such an equilibrium condition can be enforced by an equilibrium matrix $\bm{A}_0 \in \mathbb{R}^{n_{ind} \times n_{tot}}$, which comprises identity and zero matrices, i.e., 

\begin{equation}
\label{eq:18}
    \bm{A}_{0}\bm{f} = \bm{0}.
\end{equation}

The triangular lattice will exhibit the following structure for the aforementioned arrays, $\bm{f} \in \mathbb{R}^{6}$ and $\bm{A}_0 \in \mathbb{R}^{2\times6}$:

\begin{equation}
\label{eq:19}
    \bm{f} = \begin{pmatrix}
    \bm{f}_{1}\\
    \bm{f}_{2}\\
    \bm{f}_{3}
    \end{pmatrix},
    \quad
    \bm{A}_{0} = \begin{pmatrix}
    \bm{I} & \bm{I} & \bm{I}
    \end{pmatrix}.
\end{equation}

Having a closer look into $\bm{A}_{0}$ and $\bm{B}_{0}$ arrays, as mentioned in \cite{vigliotti2012linear}, it is noticed that

\begin{equation}
\label{eq:20}
    \bm{A}_{0} = \bm{B}_{0}^{T}, 
\end{equation}

and therefore, Eq. \ref{eq:18} can be rewritten as

\begin{equation}
\label{eq:21}
    \bm{B}_{0}^{T}\bm{f} = \bm{0}.
\end{equation}

Finally, the volume average of the variation of the microscale deformation work done on the unit cell $\left< \delta W \right>$, which is required for scale transition (Section \ref{subsec:2.3}), is expressed as

\begin{equation}
\label{eq:22}
    \left< \delta W \right> = \dfrac{1}{V_{uc}} \bm{f}^T \delta \bm{d},
\end{equation}

where $V_{uc}$ is the volume of the unit cell, and $\left< \cdot \right>$ denotes the volume average.

\subsection{Scale transition}
\label{subsec:2.3}

The bridging between the two scales is performed through the Hill-Mandel macrohomogeneity condition \cite{hill1963elastic,squet1985local}, based on which the volume average of the variation of the microscale deformation work done on the unit cell (Eq. \ref{eq:22}) needs to be equal to the local variation of deformation work at the macroscale (Eq. \ref{eq:5}), i.e.,

\begin{equation}
\label{eq:23}
    \delta \bar{W} = \left< \delta W \right>,
\end{equation}

\begin{equation}
\label{eq:24}
    \hat{\bar{\bm{\sigma}}}^T \delta  \hat{\bar{\bm{\varepsilon}}} = \dfrac{1}{V_{uc}} \bm{f}^T \delta \bm{d}.
\end{equation}

By taking the first derivative of the microscopic average work $\left< W \right>$ with respect to the macroscopic strain $\hat{\bar{\bm{\varepsilon}}}$, the components of the macroscopic stress  $\hat{\bar{\bm{\sigma}}}$ are computed \cite{vigliotti2012linear,vigliotti2012stiffness,vigliotti2014non} as

\begin{equation}
\label{eq:25}
    \hat{\bar{\bm{\sigma}}}^T = \dfrac{\partial{\left< W \right>}}{\partial{\hat{\bar{\bm{\varepsilon}}}}} = \dfrac{1}{V_{uc}} \bm{f}^T \dfrac{\partial{\bm{d}}}{\partial{\hat{\bar{\bm{\varepsilon}}}}}.
\end{equation}

Computing the derivative ${\partial{\bm{d}}} / {\partial{\hat{\bar{\bm{\varepsilon}}}}}$ from Eq. \ref{eq:12} as

\begin{equation}
\label{eq:26}
    \dfrac{\partial{\bm{d}}}{\partial{\hat{\bar{\bm{\varepsilon}}}}} = \bm{B}_{0} \dfrac{\partial{\bm{d}_{0}}}{\partial{\hat{\bar{\bm{\varepsilon}}}}} + \bm{B}_{e},
\end{equation}

substituting it into Eq. \ref{eq:25},

\begin{equation}
\label{eq:27}
    \hat{\bar{\bm{\sigma}}}^T = \dfrac{1}{V_{uc}} \bm{f}^T \left( \bm{B}_{0} \dfrac{\partial{\bm{d}_{0}}}{\partial{\hat{\bar{\bm{\varepsilon}}}}} + \bm{B}_{e} \right),
\end{equation}

and taking advantage of Eq. \ref{eq:21}, the array containing the macroscopic stress components $ \hat{\bar{\bm{\sigma}}}$ is simplified as

\begin{equation}
\label{eq:28}
    \hat{\bar{\bm{\sigma}}} = \dfrac{1}{V_{uc}} \left( \bm{B}_{0} \dfrac{\partial{\bm{d}_{0}}}{\partial{\hat{\bar{\bm{\varepsilon}}}}} + \bm{B}_{e} \right)^T \bm{f} = \dfrac{1}{V_{uc}} \bm{B}_{e}^T \bm{f}.
\end{equation}

Within an FE approach, the macroscopic tangent stiffness $\hat{\bar{\bm{C}}}$ is required at each macroscopic integration point (see Section \ref{sec:5}). This can be calculated by taking the derivative of the macroscopic stress field $\hat{\bar{\bm{\sigma}}}$ with respect to the macroscopic strain field $\hat{\bar{\bm{\varepsilon}}}$ as follows:

\begin{equation}
\label{eq:29}
    \hat{\bar{\bm{C}}}= \dfrac{\partial{\hat{\bar{\bm{\sigma}}}}}{\partial{\hat{\bar{\bm{\varepsilon}}}}} = \dfrac{1}{V_{uc}} \left( \left( \bm{B}_{0} \dfrac{\partial^{2}{\bm{d}_{0}}}{\partial{\hat{\bar{\bm{\varepsilon}}}^{2}}} \right)^T \bm{f} + \left( \bm{B}_{0} \dfrac{\partial{\bm{d}_{0}}}{\partial{\hat{\bar{\bm{\varepsilon}}}}} + \bm{B}_{e} \right)^T \left( \dfrac{\partial{\bm{f}}}{\partial{\bm{d}}} \dfrac{\partial{\bm{d}}}{\partial{\hat{\bar{\bm{\varepsilon}}}}} \right) \right)
\end{equation}

Recalling Eq. \ref{eq:26} and given that ${\partial^{2}{\bm{d}_{0}}} / {\partial{\hat{\bar{\bm{\varepsilon}}}^{2}}} = 0$, Eq. \ref{eq:29} is further simplified to 

\begin{equation}
\label{eq:30}
    \hat{\bar{\bm{C}}} = \dfrac{1}{V_{uc}} \left( \bm{B}_{0} \dfrac{\partial{\bm{d}_{0}}}{\partial{\hat{\bar{\bm{\varepsilon}}}}} + \bm{B}_{e} \right)^T \left( \dfrac{\partial{\bm{f}}}{\partial{\bm{d}}} \right) \left( \bm{B}_{0} \dfrac{\partial{\bm{d}_{0}}}{\partial{\hat{\bar{\bm{\varepsilon}}}}} + \bm{B}_{e} \right).
\end{equation}

Here, ${\partial{\bm{f}}} / {\partial{\bm{d}}}$ represents the stiffness matrix of the unconstrained unit cell, which will be further clarified in Section \ref{sec:3}. Subsequently, the only unknown to be computed is ${\partial{\bm{d}_{0}}} / {\partial{\hat{\bar{\bm{\varepsilon}}}}}$, which signifies the rate of change of independent DoFs with respect to the macroscopic strain. To obtain ${\partial{\bm{d}_{0}}} / {\partial{\hat{\bar{\bm{\varepsilon}}}}}$, differentiating Eq. \ref{eq:21} with respect to $\hat{\bar{\bm{\varepsilon}}}$ and inserting ${\partial{\bm{d}}} / {\partial{\hat{\bar{\bm{\varepsilon}}}}}$ from Eq. \ref{eq:26} results in

\begin{equation}
\label{eq:31}
    \bm{B}_{0}^T \dfrac{\partial{\bm{f}}}{\partial{\bm{d}}} \dfrac{\partial{\bm{d}}}{\partial{\hat{\bar{\bm{\varepsilon}}}}} = \bm{B}_{0}^T \dfrac{\partial{\bm{f}}}{\partial{\bm{d}}} \left( \bm{B}_{0} \dfrac{\partial{\bm{d}_{0}}}{\partial{\hat{\bar{\bm{\varepsilon}}}}} + \bm{B}_{e} \right) = \bm{0},
\end{equation}

which leads to the following equation after rearrangement:

\begin{equation}
\label{eq:32}
    \left( \bm{B}_{0}^T \dfrac{\partial{\bm{f}}}{\partial{\bm{d}}} \bm{B}_{0} \right) \dfrac{\partial{\bm{d}_{0}}}{\partial{\hat{\bar{\bm{\varepsilon}}}}} = - \bm{B}_{0}^T \dfrac{\partial{\bm{f}}}{\partial{\bm{d}}} \bm{B}_{e}.
\end{equation}

Finally, since the resulting matrix from $\bm{B}_{0}^T \left( {\partial{\bm{f}}} / {\partial{\bm{d}}} \right) \bm{B}_{0}$ is not invertible \cite{vigliotti2012linear,vigliotti2014non,vigliotti2012stiffness}, the Moore-Penrose pseudo inverse, denoted by $\left( \cdot \right)^+$, can be utilized to solve Eq. \ref{eq:32} for ${\partial{\bm{d}_{0}}} / {\partial{\hat{\bar{\bm{\varepsilon}}}}}$:

\begin{equation}
\label{eq:33}
    \dfrac{\partial{\bm{d}_{0}}}{\partial{\hat{\bar{\bm{\varepsilon}}}}} = - \left( \bm{B}_{0}^T \dfrac{\partial{\bm{f}}}{\partial{\bm{d}}} \bm{B}_{0} \right)^{+} \bm{B}_{0}^T \dfrac{\partial{\bm{f}}}{\partial{\bm{d}}} \bm{B}_{e}.
\end{equation}

The existence and uniqueness of the obtained solution were discussed in \cite{vigliotti2012linear}.
\section{FE formulation}
\label{sec:3}

With respect to the macroscale FE formulation, the usual procedure for a small-strain solid medium can be followed to obtain the FE system of equations. For the sake of conciseness, such established formulations are not presented here, and interested readers are referred to extensive literature, such as \cite{zienkiewicz2000finite,hughes2012finite,belytschko2014nonlinear}. In the present work, four-node quadrilateral elements with bi-linear interpolation and full integration are employed to discretize the macro-level BVP. The only difference with a single-scale FE model is that there is no explicit stress-strain constitutive relation known a priori at each macroscopic integration point, and such a relation is found by solving the microscale BVP.

Regarding the microscale, as mentioned in Section \ref{subsec:2.3}, the stiffness matrix of the unconstrained unit cell $\bm{K}_{uc} \in \mathbb{R}^{n_{tot} \times n_{tot}}$ is expressed by

\begin{equation}
\label{eq:34}
    \bm{K}_{uc} = \dfrac{\partial{\bm{f}}}{\partial{\bm{d}}}.
\end{equation}

Such a stiffness matrix can be assembled from the element stiffness matrices of all the lattice struts $\bm{K}_{glo}^{e} \in \mathbb{R}^{2N_{dim} \times 2N_{dim}}$ \cite{fish2007first} through

\begin{equation}
\label{eq:35}
    \bm{K}_{uc} = \sum_{e=1}^{n_e} {\bm{\Lambda}^{e}}^{^T} \bm{K}_{glo}^{e} \; \bm{\Lambda}^{e},
\end{equation}

where $n_e$ is the number of individual struts of the unit cell, and $\bm{\Lambda}^{e} \in \mathbb{R}^{2N_{dim} \times n_{tot}}$ represents the element gather-scatter Boolean matrix \cite{fish2007first}, connecting the element DoFs to the global ones. The element stiffness matrix in the global coordinate system $\bm{K}_{glo}^{e}$ is related to its counterpart in the local coordinate system $\bm{K}_{loc}^{e} \in \mathbb{R}^{2\times2}$ \cite{ochsner2014} as 

\begin{equation}
\label{eq:36}
    \bm{K}_{glo}^{e} = {\bm{T}^{e}}^{^T} \bm{K}_{loc}^{e} \; \bm{T}^{e}.
\end{equation}

\begin{figure}[ht]
    \centering
    \includegraphics[keepaspectratio=true, width=0.4\linewidth]{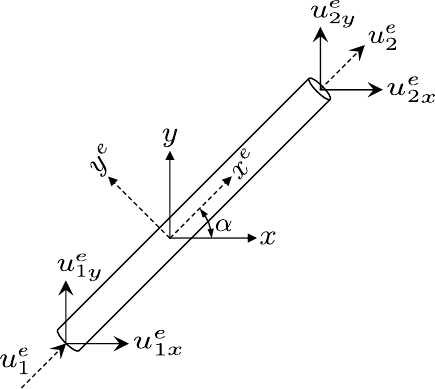}
    \caption{Schematic representation of a truss element in the global and local coordinate systems (i.e., $x$-$y$ and $x^{e}$-$y^{e}$ coordinate systems, respectively). $u_{1x}^{e}$, $u_{1y}^{e}$, $u_{2x}^{e}$ and $u_{2y}^{e}$ denote the nodal displacements in the global coordinate system, while $u_{1}^{e}$ and $u_{2}^{e}$ show such displacements in the local coordinate system.}
    \label{fig:3}
\end{figure}

Here, $\bm{T}^{e} \in \mathbb{R}^{2 \times 2N_{dim}}$ is a transformation matrix, transforming the quantities from the element local coordinate system to the global one and vice versa (Figure \ref{fig:3}). For the 2D case in the $x-y$ plane, such a transformation matrix can be expressed in terms of the angle between the local and global coordinate systems $\alpha$ as

\begin{equation}
\label{eq:37}
    \bm{T}^{e} = \begin{pmatrix*}[r]
    \cos{\alpha} & \sin{\alpha} & 0 & 0 \\
    0 & 0 & \cos{\alpha} & \sin{\alpha}
    \end{pmatrix*}.
\end{equation}

Following any FE approach (e.g., variational, energy, or weighted residual approaches) \cite{ochsner2013one,ochsner2014}, the element stiffness matrix $\bm{K}_{loc}^{e}$ is obtained by evaluating the following integral:

\begin{equation}
\label{eq:38}
    \bm{K}_{loc}^{e} = \int_\Omega {\dfrac{\mathrm{d}{\bm{N}^e}}{\mathrm{d}{x^e}}}^T D^{e} \; {\dfrac{\mathrm{d}{\bm{N}^e}}{\mathrm{d}{x^e}}} \; \mathrm{d} \Omega.
\end{equation}

In the above, $D^e$ denotes the tangent modulus, which is equal to the elastic modulus $E^e$ below the proportional limit (i.e., in the elastic regime) and is replaced by the elastoplastic tangent modulus $C_{ep}^{e}$ beyond plastic yielding (see Sections \ref{sec:4} and \ref{sec:5} for more details). Furthermore, $\bm{N}^e \in \mathbb{R}^{1\times2}$ represents the matrix of element shape functions, $N_{1}^{e} \left(x^{e}\right)$ and $N_{2}^{e} \left(x^{e}\right)$, which can be defined as follows if linear interpolation along the element length $L^e$ is considered:

\begin{equation}
\label{eq:39}
    \bm{N}^{e} = \begin{pmatrix}
         N_{1}^{e} \left(x^{e}\right) &  N_{2}^{e} \left(x^{e}\right)
    \end{pmatrix},
    \quad
    N_{1}^{e} \left(x^{e}\right) = \dfrac{1}{L^e} \left( x_{2}^{e} - x^{e} \right),
    \quad
    N_{2}^{e} \left(x^{e}\right) = \dfrac{1}{L^e} \left( x^{e} - x_{1}^{e} \right).
\end{equation}

Assuming constant cross section $A^e$ and tangent modulus $D^e$ along the element length $L^e$, the element stiffness matrix will take the final form of

\begin{equation}
\label{eq:40}
    \bm{K}_{loc}^{e} = \dfrac{A^e D^e}{L^e}\begin{pmatrix*}[r]
    1 & -1 \\
    -1 & 1
    \end{pmatrix*}.
\end{equation}

Now, it remains to compute the axial strain of each strut $\varepsilon^e$, which is required to define the constitutive behavior of the microscale unit cell (Section \ref{sec:4}). To do so, the element nodal displacements in the global coordinate system $\bm{u}_{glo}^{e} \in \mathbb{R}^{2 N_{dim}}$ can be extracted from the nodal DoFs of the unit cell $\bm{d}$ as

\begin{equation}
\label{eq:41}
    \bm{u}_{glo}^{e} = {\bm{\Lambda}^{e}} \bm{d}.
\end{equation}

Then, as depicted in Figure \ref{fig:3}, the element nodal displacements can be transformed from the global coordinate system (i.e., $\bm{u}_{glo}^{e}$) to the local one (i.e., $\bm{u}_{loc}^{e} \in \mathbb{R}^2$) by

\begin{equation}
\label{eq:42}
    \bm{u}_{loc}^{e} = \bm{T} \bm{u}_{glo}^{e},
\end{equation}

where, for the present 2D case,

\begin{equation}
\label{eq:43}
    \bm{u}_{loc}^{e} = \begin{pmatrix}
    {u}_{1}^{e}\\
    {u}_{2}^{e}
    \end{pmatrix},
    \quad
    \bm{u}_{glo}^{e} = \begin{pmatrix}
    {u}_{1x}^{e}\\
    {u}_{1y}^{e}\\
    {u}_{2x}^{e}\\
    {u}_{2y}^{e}
    \end{pmatrix}.
\end{equation}

Eventually, the axial displacements in the element local coordinate system, ${u}_{1}^{e}$ and ${u}_{2}^{e}$, are used to compute the element axial strain $\varepsilon^e$ as

\begin{equation}
\label{eq:44}
    \varepsilon^e = \dfrac{u_{2}^{e} - u_{1}^{e}} {L^e}.
\end{equation}
\section{Constitutive model}
\label{sec:4}

In the two-scale computational homogenization method, the stress-strain constitutive relation is defined at the microscale. In order to establish the relationship between the axial stress of each lattice strut $\sigma^e$ and its corresponding axial strain $\varepsilon^e$, a 1D elastoplastic material behavior with mixed nonlinear Voce isotropic hardening \cite{voce1955practical} and linear kinematic hardening \cite{prager1945strain,ziegler1959modification} is considered \cite{simo2006computational}. For readers' convenience, such a well-established material law is briefly revisited in this section. For notational clarity, the superscript $\left( \cdot \right)^e$ is dropped hereinafter. To start with, the total strain $\varepsilon$ is decomposed into an elastic part $\varepsilon^{el}$ and a plastic contribution $\varepsilon^{pl}$ \cite{simo2006computational}: 

\begin{equation}
\label{eq:45}
    \varepsilon = \varepsilon^{el} + \varepsilon^{pl}.
\end{equation}

The stress field $\sigma$ is associated only with the elastic part of strain $\varepsilon^{el}$ by the elastic modulus $E$ as

\begin{equation}
\label{eq:46}
    \sigma = E \varepsilon^{el} = E \left( \varepsilon - \varepsilon^{pl} \right).
\end{equation}

In order for the stress $\sigma$ to be admissible, the following yield condition needs to be fulfilled:

\begin{equation}
\label{eq:47}
    \Phi(\sigma, q, \alpha) = \left| \sigma - q \right| - \sigma_{y}(\alpha) \leq 0.
\end{equation}

Here, $\Phi(\sigma, q, \alpha)$ denotes the yield function, $q$ represents the back stress, controlling the location of the center of the yield surface, and $\sigma_{y}(\alpha)$ is the isotropic hardening function, governing the expansion of the yield surface based on the evolution of the internal hardening variable $\alpha$. The plastic strain rate $\dot{\varepsilon}^{pl}$, where $\dot{\left( \cdot \right)} = {\partial{\left( \cdot \right)}} / {\partial{t}}$, can be found by using the associative flow rule as

\begin{equation}
\label{eq:48}
    \dot{\varepsilon}^{pl} = \dot{\lambda} \dfrac{\partial{\Phi}}{\partial{\sigma}} = \dot{\lambda} \sgn{(\sigma-q)}.
\end{equation}

In the above, $\dot{\lambda} \geq 0$ is the plastic multiplier, and $\sgn \left(x\right)$ denotes the signum function, returning the sign of a real number $x$. Considering linear kinematic hardening, the back stress rate $\dot{q}$ is expressed by Ziegler's rule as 

\begin{equation}
\label{eq:49}
    \dot{q} = H \dot{\varepsilon}^{pl} = H \dot{\lambda} \sgn{(\sigma-q)},
\end{equation}

where $H$ is the kinematic hardening modulus. The rate of the internal hardening parameter $\dot{\alpha}$ is expressed as a non-negative function of the amount of plastic flow, i.e., $\dot{\alpha} = \left| \dot{\varepsilon}^{pl} \right| = \dot{\lambda}$. Then, considering the nonlinear Voce isotropic hardening \cite{voce1955practical} 

\begin{equation}
\label{eq:55}
    \sigma_{y} = \sigma_{y_0} + Q_{\infty} \left( 1 - \mathrm{e}^{- b \alpha} \right),
\end{equation}

and taking advantage of the Kuhn-Tucker loading\fshyp{}unloading conditions (i.e., $\dot{\lambda} \geq 0, \; \Phi \leq 0, \; \dot{\lambda} \Phi = 0$) along with the consistency condition (i.e., $\dot{\lambda} \dot{\Phi} = 0 \; \text{if } \Phi = 0$), the plastic multiplier $\dot{\lambda}$ can be obtained as

\begin{equation}
\label{eq:56}
    \dot{\lambda} = \dfrac{E \sgn{(\sigma-q)}}{E + H + Q_{\infty} b \, \mathrm{e}^{- b \alpha}} \dot{\varepsilon},
\end{equation}

with $\sigma_{y_0}$ being the initial yield stress and $Q_{\infty}$ and $b$ denoting the saturation flow stress and saturation exponent, respectively. Finally, the relation between the stress rate $\dot{\sigma}$ and the strain rate $\dot{\varepsilon}$ can be established as

\begin{equation}
\label{eq:59}
    \dot{\sigma} = 
    \begin{cases}
        E \dot{\varepsilon},& \text{if } \dot{\lambda}=0 \\
        C_{ep} \dot{\varepsilon},& \text{if } \dot{\lambda}>0
    \end{cases}
     \quad \text{with } C_{ep} =  \dfrac{E \left( H + Q_{\infty} b \, \mathrm{e}^{- b \alpha} \right)}{E + H + Q_{\infty} b \, \mathrm{e}^{- b \alpha}}.
\end{equation}

In the above, $C_{ep}$ is the elastoplastic modulus, which replaces the elastic modules $E$ during plastic flow (i.e., when $\dot{\lambda} > 0$).
\section{Algorithmic implementation}
\label{sec:5}

The solution procedure for the employed two-scale homogenization scheme includes solving the macro- and microscale BVPs in a concurrent setting with on-the-fly information exchange between the two scales (e.g., see \cite{kouznetsova2001approach,kouznetsova2004computational}). After setting up the macroscopic geometry, discretizing the domain with a finite number of elements and applying BCs, a standard iterative solution procedure for nonlinear FE problems starts. An increment of the applied macroscopic load (or displacement in the case of displacement control) is applied, and the computed macroscopic strain $\bar{\bm{\varepsilon}}$ at each integration point of the macroscopic domain is transferred to the microscale, where it governs the deformation of the unit cell. In the microscale, the macroscopic strain field $\bar{\bm{\varepsilon}}$ is used to compute the nodal displacements of the unit cell $\bm{d}$ and, consequently, the axial strain of each strut $\varepsilon$. Subsequently, the well-established elastoplastic return mapping algorithm (e.g., see \cite{simo2006computational,Neto2011computational}) is utilized to compute the algorithmic tangent modulus $D^e$ and, thereby, the element stiffness matrices $\bm{K}_{loc}^{e}$ and $\bm{K}_{glo}^{e}$ for each strut. Then, after assembling the stiffness matrix of the unit cell $\bm{K}_{uc}$, the macroscopic stress $\bar{\bm{\sigma}}$ and tangent stiffness $\bar{\bm{C}}$ are computed and transferred back to the macroscale. Afterwards, due to the availability of stress $\bar{\bm{\sigma}}$ at each integration point of the macroscopic body, the macroscopic internal forces can be computed. If these internal forces are in equilibrium with the macroscopic applied loads (or the resulting reaction forces in the case of displacement control), incremental convergence is accomplished, and the next macroscopic load increment can be applied. On the other hand, if convergence is not achieved, the already available macroscopic tangent $\bar{\bm{C}}$ at each macroscopic integration point, obtained from microscopic unit cell calculations, is used to assemble the macroscopic stiffness matrix. Then, by solving the macroscopic BVP with the new stiffness matrix, an updated estimation of the macroscopic displacement field $\bar{\bm{u}}$ and, as a result, an updated macroscopic strain field $\bar{\bm{\varepsilon}}$, which will be transferred to the microscale, are obtained. This iterative procedure is continued until incremental convergence at the macroscale is achieved.

It is worth noting that before imposing the first load increment, an initialization is necessary to compute the initial macroscopic tangent $\bar{\bm{C}}$ at each macroscopic integration point \cite{kouznetsova2004computational}. During such an initialization step, the deformation across the entire macroscopic domain is assumed to be zero (i.e., $\bar{\bm{\varepsilon}}=\bm{0}$). This results in obtaining the initial macroscopic tangent $\bar{\bm{C}}$ from the microscopic computations of an undeformed unit cell.

The algorithmic summary of this two-scale solution procedure with on-the-fly information exchange is presented in Table \ref{tab:1}.

\begin{table}[h!]
\caption{Algorithmic summary of the two-scale incremental-iterative solution method with on-the-fly information exchange for lattice structures with elastoplastic material formulation.}
\label{tab:1}
\centering
    \includegraphics[keepaspectratio=true, width=\linewidth]{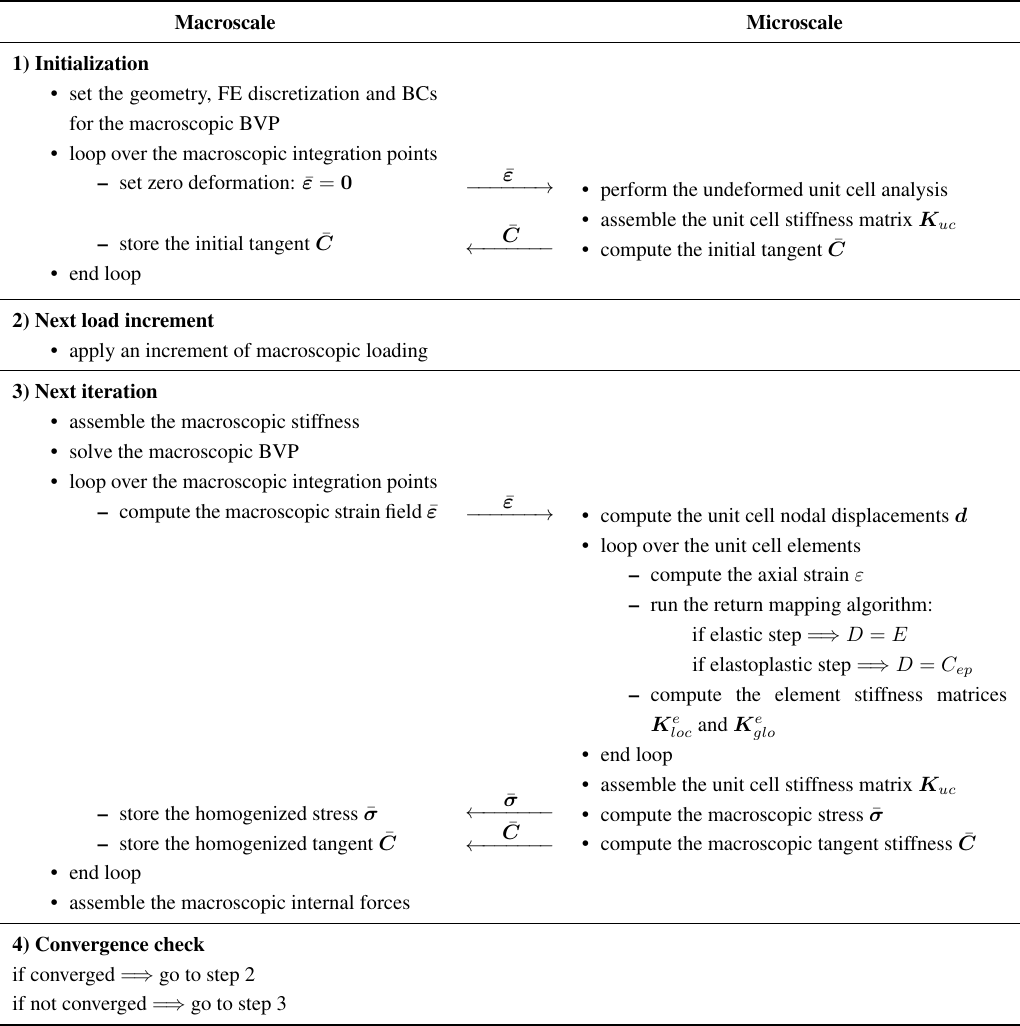}
\end{table}

\section{Numerical studies}
\label{sec:6}

In this section, a number of numerical examples are provided to investigate the performance of the employed homogenization technique from different aspects. The open source FE code PyFEM \cite{de2012nonlinear} is used to implement the two-scale homogenization scheme. Four-node bi-linear quadrilateral elements are used to discretize the macroscale structure, and the microscale lattice structures are discretized by two-node linear truss elements. In addition, in order to provide reference solutions for comparisons, direct numerical simulations (DNS), i.e., simulations considering full-scale discrete models of the structures, are performed by the commercial FE software Abaqus using two-node linear displacement truss elements (i.e., T2D2 elements in Abaqus naming convention). The material parameters of the additively manufactured AlSi10Mg are used from the reference \cite{servatan2023ratcheting} as $E = 70000 \, \mathrm{MPa}$, $H = 16000 \, \mathrm{MPa}$, $\sigma_{y_0} = 190 \, \mathrm{MPa}$, $Q_{\infty} = 90 \, \mathrm{MPa}$ and $b = 13.5$. Unless otherwise specified, the unit cell side length and the cross-sectional area of each strut are assumed to be $L_{uc} = 1 \, \mathrm{mm}$ and $A = 0.1 \, \mathrm{mm}^2$, respectively, in the subsequent examples.

It is to be noted that based on the lattice topology and BCs, a group of lattice struts might undergo compression during loading. In such a scenario, the possible buckling of each strut needs to be taken into account as well. However, in the present work, in order to solely focus on elastoplasticity, the dimensions of the lattice struts are chosen in such a way that buckling is prevented within the range of the applied loadings. Even so, the current framework is flexible enough to accommodate buckling, which will be briefly discussed in Section \ref{sec:7} as an outlook for future works. 

\subsection{Double-clamped beam with central loading}
\label{subsec:6.1}

\begin{figure}[h]
    \centering
    \includegraphics[keepaspectratio=true]{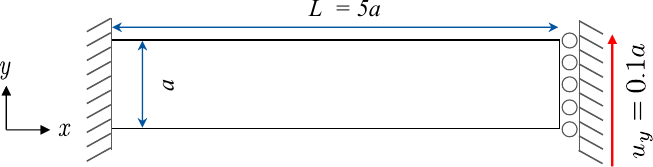}
    \caption{Geometry and BCs for a double-clamped beam, which, due to symmetry, is depicted as a one-half model with a length of $L$ and a thickness of $a$. The left boundary is fixed in both the $x$ and $y$ directions; the right boundary is fixed only in the $x$ direction, and a vertical displacement of $u_y = 0.1a$ is applied to the right boundary.}
    \label{fig:4}
\end{figure}

As the first numerical example, a centrally loaded double-clamped beam is considered. Due to symmetry, it is sufficient to study a one-half model as presented in Figure \ref{fig:4}. The left side of the beam is fixed in both the $x$ and $y$ directions, and the right edge is fixed only in the $x$ direction. A vertical displacement of $u_y = 0.1a$ is applied to the right boundary.

\begin{figure}[h]
    \centering
    \includegraphics[keepaspectratio=true]{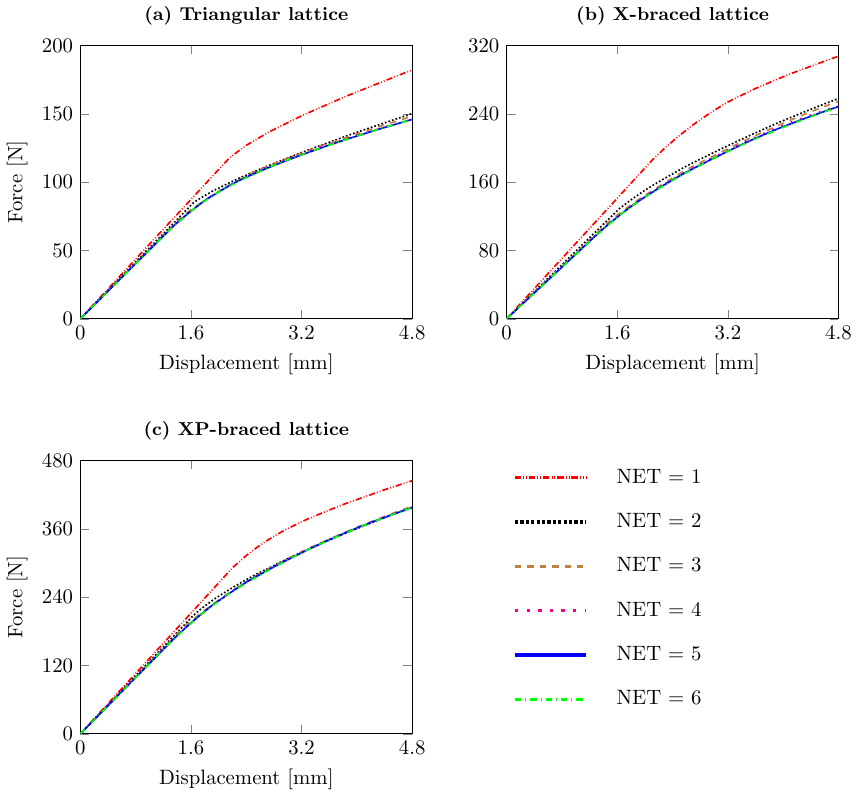}
    \caption{Force-displacement curves of the double-clamped beam with different numbers of elements through the thickness for the \textbf{a)} triangular lattice, \textbf{b)} X-braced lattice and \textbf{c)} XP-braced lattice.}
    \label{fig:5}
\end{figure}

First of all, it is considered that the beam is composed of $240\times48$ lattices, and the mesh convergence behavior of the proposed homogenization method is studied by solving the macroscopic problem for different mesh distributions. 30 elements are considered along the beam length, and the number of elements in the thickness (NET) is varied from 1 to 6 elements. The sum of forces in the $y$ direction on the right boundary vs. its displacement is plotted for the triangular, X-braced and XP-braced lattices and different NET in Figure \ref{fig:5}. It is observed that for all three lattice topologies, the force-displacement curves are converged by using 5 and more elements along the beam thickness (i.e., NET = 5 and NET = 6). However, to ensure the element aspect ratio of 1, the mesh distribution of $30\times6$ is used in the following.

\begin{figure}[h]
    \centering
    \includegraphics[keepaspectratio=true]{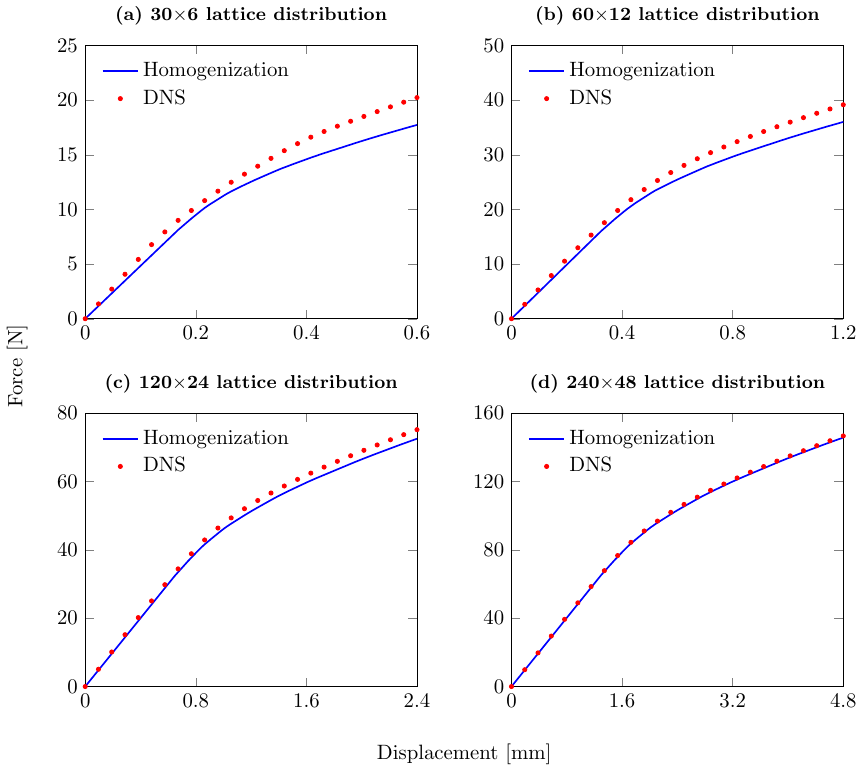}
    \caption{Force-displacement curves of the double-clamped beam for the triangular lattice with different lattice distributions, including \textbf{a)} $30\times6$, \textbf{b)} $60\times12$, \textbf{c)} $120\times24$ and \textbf{d)} $240\times48$ lattices.}
    \label{fig:6}
\end{figure}

\begin{figure}[h]
    \centering
    \includegraphics[keepaspectratio=true]{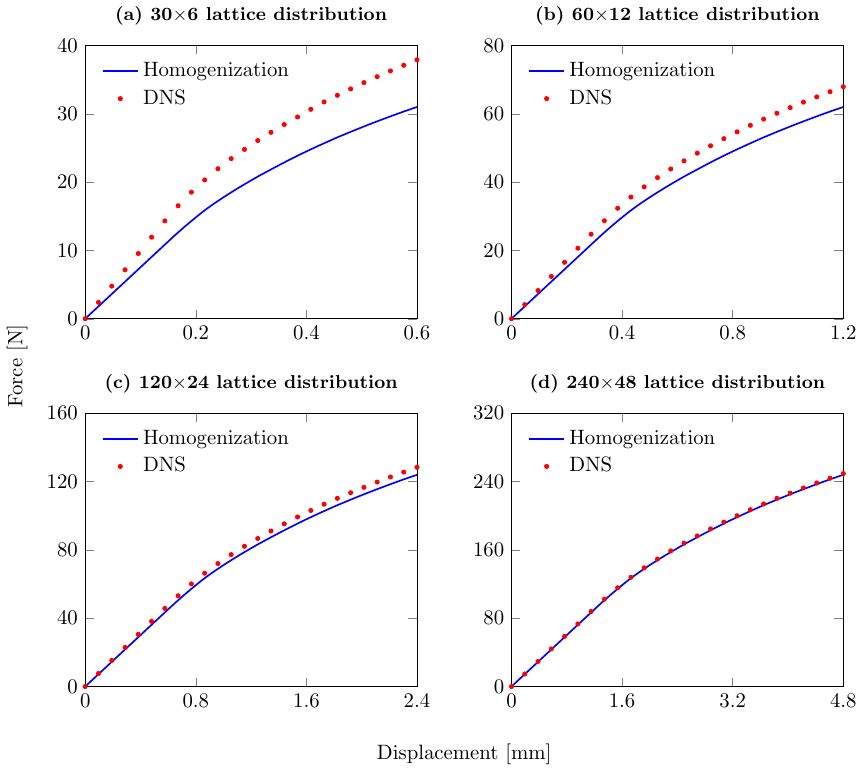}
    \caption{Force-displacement curves of the double-clamped beam for the X-braced lattice with different lattice distributions, including \textbf{a)} $30\times6$, \textbf{b)} $60\times12$, \textbf{c)} $120\times24$ and \textbf{d)} $240\times48$ lattices.}
    \label{fig:7}
\end{figure}

\begin{figure}[h]
    \centering
    \includegraphics[keepaspectratio=true]{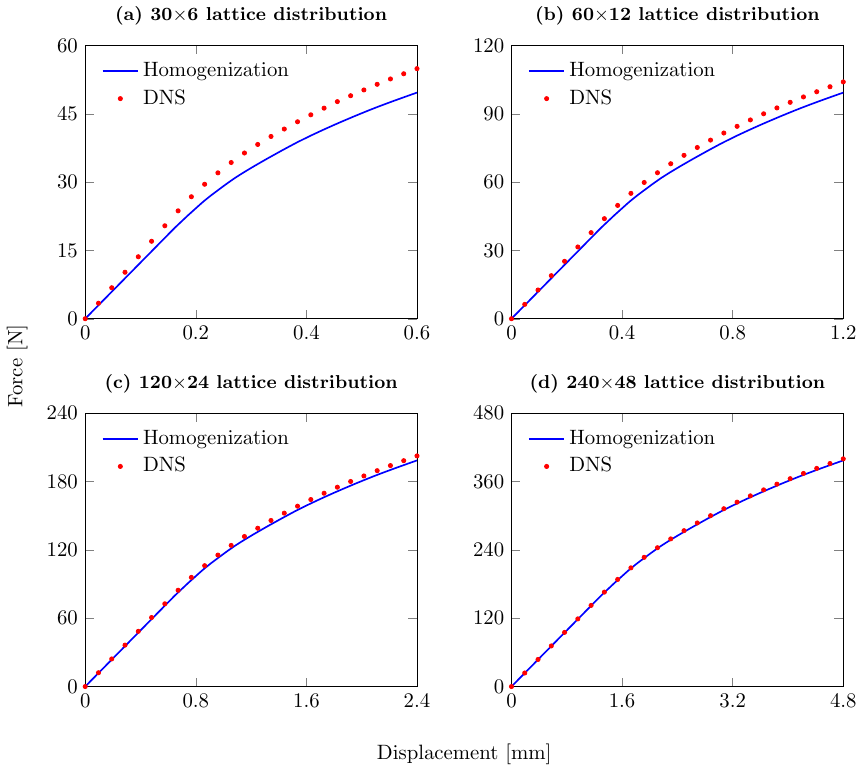}
    \caption{Force-displacement curves of the double-clamped beam for the XP-braced lattice with different lattice distributions, including \textbf{a)} $30\times6$, \textbf{b)} $60\times12$, \textbf{c)} $120\times24$ and \textbf{d)} $240\times48$ lattices.}
    \label{fig:8}
\end{figure}

In the next step, the fulfillment of the principle of scale separation for the current homogenization approach is examined. Based on this principle, the microscopic length scale (i.e., the lattice size) is supposed to be significantly smaller than the characteristic length of the macroscopic problem (i.e., the size of the macroscale structure) to obtain realistic homogenization \cite{geers2010multi}. In order to investigate this question, it is considered that the macroscale beam is composed of different lattice distributions, including $30\times6$, $60\times12$, $120\times24$ and $240\times48$ lattices. More precisely, considering that the lattice side length (i.e., the microscopic length scale) remains constant and equal to $L_{uc} = 1 \, \mathrm{mm}$, the macroscopic beam length (i.e., the macroscopic length scale) will change from $30 \, \mathrm{mm}$ to $240 \, \mathrm{mm}$ by using the aforementioned lattice distributions, respectively. The force-displacement curves for the mentioned distributions of the triangular, X-braced and XP-braced lattices are presented in Figures \ref{fig:6}-\ref{fig:8}, respectively. It can be seen that for all three lattice families, by increasing the number of lattices from $30\times6$ to $240\times48$ (i.e., increasing the macroscopic characteristic length, while its microscopic counterpart remains constant), the agreement between the results obtained from homogenization and DNS is also increased, insofar as a close agreement is inspected for the $240\times48$ lattice distribution. It is also observed that for all the lattice topologies and all the lattice distributions, the disagreement between the homogenization results and those of DNS is increased by crossing the proportional limit, and such a discrepancy becomes more significant as the structure undergoes more plastic deformation. Finally, it can be concluded that by using a sufficiently large number of lattices (e.g., a minimum of $240\times48$ in this specific example), the employed homogenization scheme results in adequately precise solutions, which are in agreement with full-scale discrete simulations not only in the elastic regime but also beyond the proportional limit. 

\subsection{Plate under tension}
\label{subsec:6.2}

\begin{figure}[h]
    \centering
    \includegraphics[keepaspectratio=true, width=0.75\linewidth]{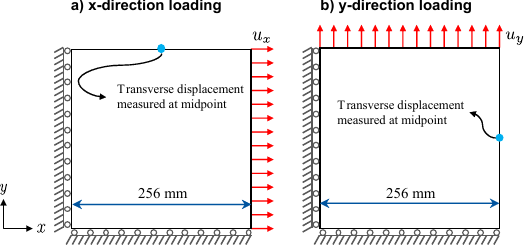}
    \caption{Geometry and BCs for a square plate with a side length of $256 \, \mathrm{mm}$ under tension. The left boundary is fixed in the $x$ direction; the bottom boundary is fixed in the $y$ direction, and a tensile displacement is applied to the right or top boundaries, referred to as \textbf{a)} $x$-direction loading and \textbf{b)} $y$-direction loading, respectively. The transverse displacement is measured at the midpoint of the lateral edge.} 
    \label{fig:9}
\end{figure}

As the next numerical example, a square plate with a side length of $256 \, \mathrm{mm}$ under tensile loading is considered. As depicted in Figure \ref{fig:9}a, the left and bottom edges are fixed in the $x$ and $y$ directions, respectively, and a tensile displacement of $u_x = 1 \, \mathrm{mm}$ is applied on the right boundary. It is to be noted that the X-braced and XP-braced lattices possess a discrete rotational symmetry of 4th order, meaning that the original topology is obtained by a $90^{\circ}$ rotation. As a result, loading in the $x$ and $y$ directions will lead to identical results. However, this does not hold for the triangular lattice, meaning that the $x$- and $y$-direction loadings will give rise to different behaviors. Therefore, a second case considering the $y$-direction loading of $u_y = 1 \, \mathrm{mm}$ (Figure \ref{fig:9}b) is examined solely for the triangular lattice. Considering that the lattice side length is $L_{uc}= 1 \, \mathrm{mm}$, full-scale simulations (i.e., DNS) are also performed for a $256\times256$ lattice distribution. 

\begin{figure}[h]
    \centering
    \includegraphics[keepaspectratio=true]{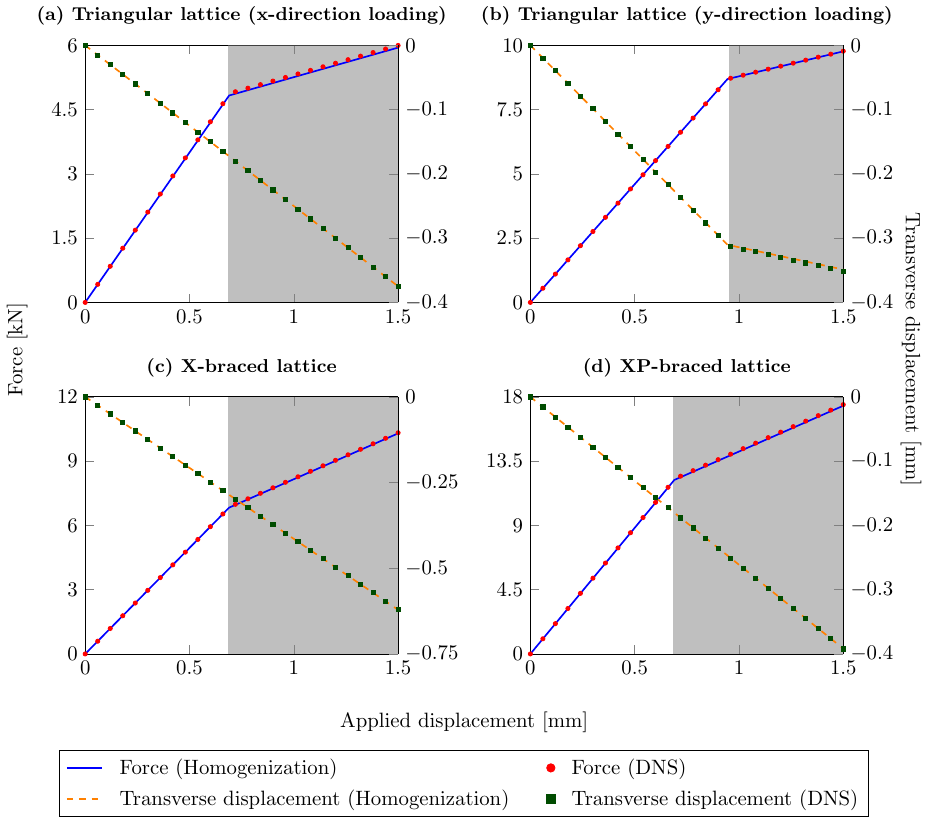}
    \caption{Sum of forces corresponding to the boundary with applied displacement (left $y$ axis) and the measured transverse displacement at the midpoint of the lateral edge (right $y$ axis) plotted vs. the applied displacement ($x$ axis) for the plate under tension with the \textbf{a)} triangular lattice ($x$-direction loading), \textbf{b)} triangular lattice ($y$-direction loading), \textbf{c)} X-braced lattice and \textbf{d)} XP-braced lattice.}
    \label{fig:10}
\end{figure}

In this example, in addition to the sum of forces, the transverse displacement at the midpoint of the lateral edge (as depicted in Figure \ref{fig:9}) is also measured and plotted vs. the applied displacement in Figure \ref{fig:10} for all the considered unit cell topologies. It is evident that the employed homogenization approach follows a remarkably similar force-displacement path as that of DNS, and it also quite precisely mimics the transverse displacement throughout the deformation.

\begin{figure}[h]
    \centering
    \includegraphics[keepaspectratio=true,width=0.6\textwidth]{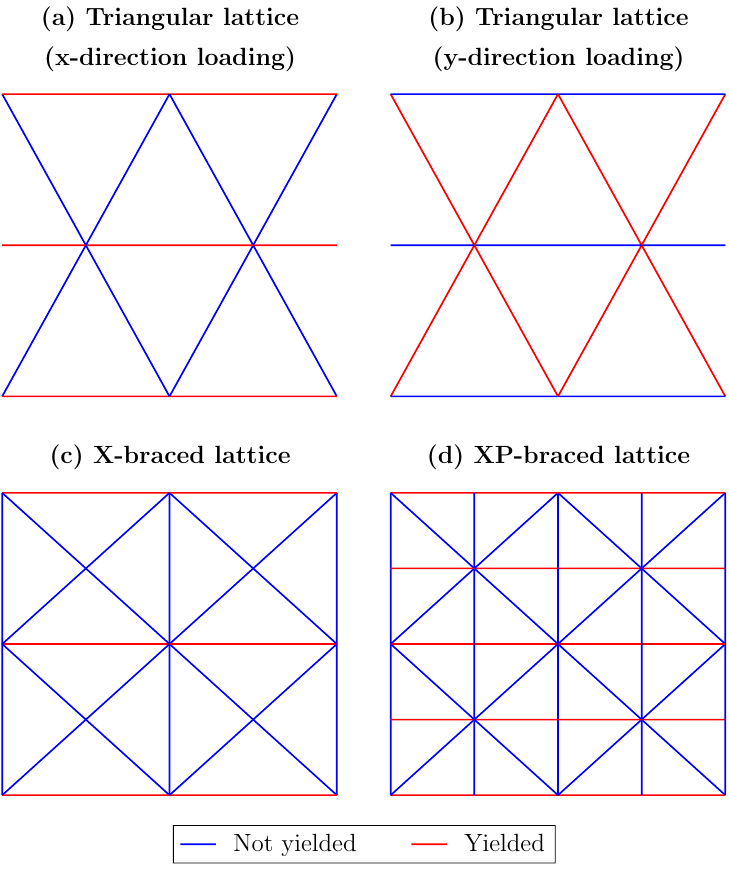}
    \caption{Illustration of the yielded and not yielded struts within a $2\times2$ lattice distribution extracted from an arbitrary position of the plate under tension modeled with DNS for the \textbf{a)} triangular lattice ($x$-direction loading), \textbf{b)} triangular lattice ($y$-direction loading), \textbf{c)} X-braced lattice and \textbf{d)} XP-braced lattice.}
    \label{fig:11}
\end{figure}

An interesting feature observed in this example is the rate of change of transverse displacement with respect to the applied displacement for the triangular lattice under tension in the $y$ direction (i.e., Figure \ref{fig:10}b). More specifically, it is seen that the triangular lattice under $x$-direction loading as well as the X-braced and XP-braced unit cells exhibit a constant rate of change of lateral displacement throughout the loading path, both below and beyond the proportional limit (i.e., the white and gray zones in Figure \ref{fig:10}). However, for the triangular lattice under $y$-direction loading, as soon as the proportional limit is crossed and the plastic deformation starts, the transverse displacement develops at a different rate. This can be further explained by a closer look into the behavior of the individual struts in the direction transverse to the applied displacement. To better demonstrate, an illustration of the yielded and not yielded struts within a $2\times2$ lattice distribution extracted from an arbitrary position in the plate under tension (i.e., from the entire $256\times256$ distribution modeled with DNS) is presented in Figure \ref{fig:11} for all four cases. It is an arbitrary $2\times2$ lattice distribution because the pattern for the yielded elements is homogeneous throughout the entire plate, and any unit cell within the plate exhibits the same yielded struts. As can be seen, in the three cases with a constant rate of change of transverse displacement (Figures \ref{fig:11}a, c and d), the struts contributing to the lateral deformation (i.e., the $y$ direction) do not yield and continue to behave elastically, while the struts straightly aligned in the direction of the applied displacement (i.e., the horizontal struts in the $x$ direction) cross the proportional limit and cause the elastoplastic response for the overall structural behavior (i.e., the force-displacement curves in Figure \ref{fig:10}). As a result, the lateral displacement follows a curve with a constant slope resulting from the elastic deformation of the struts involved in the displacement of the lateral edge. This is while for the triangular lattice with the applied displacement in the $y$ direction (Figure \ref{eq:11}b), the legs of the triangles (i.e., the oblique struts), which contribute to both the longitudinal and transverse deformations, also yield, resulting in a change in the slope of the transverse displacement curve beyond the proportional limit.

Given the measured transverse displacement, it is straightforward to compute the effective elastic Poisson's ratio for each of the lattices by using the definition

\begin{equation}
\label{eq:73}
    \overline{\nu} = - \dfrac{\varepsilon_{trans}}{\varepsilon_{axial}} \qquad \mathrm{with} \quad \varepsilon_{trans} = \dfrac{u_{trans}}{L_{trans}}, \quad \varepsilon_{axial} = \dfrac{u_{axial}}{L_{axial}},
\end{equation}

where the displacements $u_{axial}$ and $u_{trans}$, and the side lengths $L_{axial}$ and $L_{trans}$ denote the displacements and lengths of the longitudinal and transverse boundaries, respectively. As the considered plate has a square-shaped initial geometry (i.e., $L_{axial} = L_{trans}$), the definition of the effective elastic Poisson's ratio can be further simplified as

\begin{equation}
\label{eq:74}
    \overline{\nu} = - \dfrac{u_{trans}}{u_{axial}}.
\end{equation}

A noteworthy point to be made here is that due to the $90^{\circ}$ rotational symmetry of the X-braced and XP-braced lattices, the effective Poisson's ratios in both the $x$ and $y$ directions are the same, i.e., $\overline{\nu}_{xy} = \overline{\nu}_{yx}$. On the other hand, due to the lack of the $90^{\circ}$ rotational symmetry for the triangular lattice, the effective Poisson's ratio in each direction is different (i.e., $\overline{\nu}_{xy} \neq \overline{\nu}_{yx}$) and needs to be computed separately.

The computed effective elastic Poisson's ratios for all four cases are presented in Table \ref{tab:3}, where identical solutions (up to 3 decimal places) for the homogenization approach and DNS are obtained for the triangular lattice under $x$-direction loading as well as the X-braced and XP-braced unit cells. However, for the triangular lattice under $y$-direction loading, a negligible error of approximately $0.3 \%$ is observed. With respect to the behavior of each lattice topology, the triangular lattice, when loaded in the $x$ direction, has the lowest effective elastic Poisson's ratio, while the X-braced lattice shows the highest value. It is also observed that the XP-braced unit cell results in an effective elastic Poisson's ratio relatively similar to that of the triangular unit cell with $x$-direction loading.

\begin{table}[h]
\caption{Effective elastic Poisson's ratios for different lattice topologies computed from DNS and homogenization for the plate under tension.}
\label{tab:3}
\resizebox{\columnwidth}{!}{%
\begin{tabular}{@{}lcccc@{}}
\toprule
Model          & Triangle ($x$-direction loading)            & Triangle ($y$-direction loading)            & X-braced          & XP-braced          \\ \midrule
DNS            & 0.250                                     & 0.328                                     & 0.414             & 0.261              \\
Homogenization & 0.250                                     & 0.329                                     & 0.414             & 0.261              \\ \bottomrule
\end{tabular}%
}
\end{table}

As discussed in Section \ref{subsec:6.1}, in order for the homogenization approach to provide precise enough solutions, the two scales must be adequately separated (i.e., the separation of scales must hold). Once again, this principle is investigated for the present numerical example, but this time instead of forces, a displacement-related quantity such as Poisson's ratio is considered for investigation. To do so, the case with the variable rate of change of transverse displacement, i.e., the triangular lattice with $y$-direction loading, is considered and solved for different distributions of unit cells, from $16\times16$ to $256\times256$. Here, the same definition of Eqs. \ref{eq:73} and \ref{eq:74} is used to compute Poisson's ratio in both the elastic and elastoplastic regimes. Therefore, in order to differentiate from the effective elastic Poisson's ratio, which is only defined in the elastic region, it is referred to as the effective elastoplastic Poisson's ratio.

\begin{figure}[h]
    \centering
    \includegraphics[keepaspectratio=true]{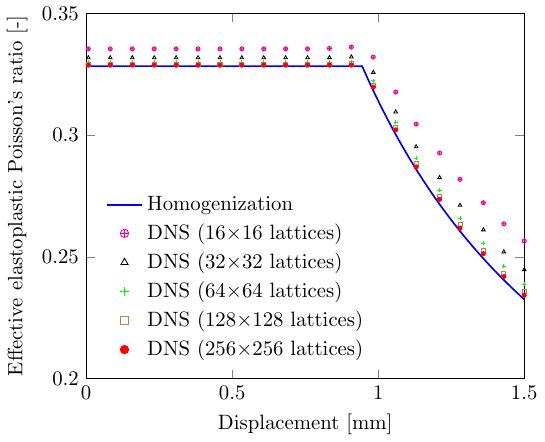}
    \caption{Effective elastoplastic Poisson's ratio for the triangular lattice with $y$-direction loading computed from homogenization and DNS for the plate under tension with different lattice distributions.}
    \label{fig:12}
\end{figure}

Figure \ref{fig:12} demonstrates the plots of the effective elastoplastic Poisson's ratio vs. the applied displacement (in the $y$ direction) for different distributions of the triangular lattice, from $16\times16$ to $256\times256$. For different lattice distributions, the side length of the unit cell is considered in such a way that the overall size of the plate always remains constant and equal to $256\times256 \, \mathrm{mm}$. For instance, for the $16\times16$ lattice distribution, the unit cell side length is considered to be $L_{uc} = 16 \, \mathrm{mm}$; for $32\times32$ lattices, the lattice side length is $L_{uc} = 8 \, \mathrm{mm}$, and so forth. In this case, the homogenization approach yields the same displacement field and, consequently, the same effective Poisson's ratio for different lattice distributions, which helps to reduce the number of figures. As can be seen in Figure \ref{fig:12}, by increasing the number of lattices, the performance of the homogenization scheme is also enhanced, so that when an adequate number of lattices are used (e.g., $256\times256$), a very good agreement between the homogenization and DNS results is witnessed. It is also noticed that the performance of the homogenization technique is slightly more pronounced in the elastic region, and the small gap between the results becomes more apparent as the structure experiences more plastic deformation.

\subsection{Symmetrically notched dog-bone specimen}
\label{subsec:6.3}

\begin{figure}[h]
    \centering
    \includegraphics[keepaspectratio=true]{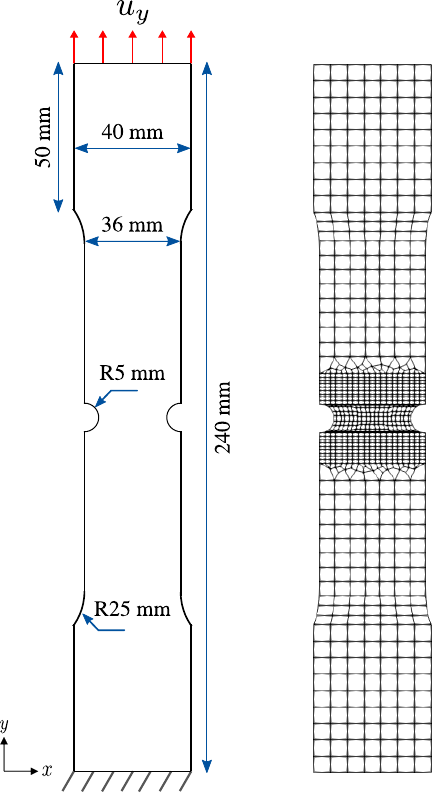}
    \caption{Geometry, BCs and FE discretization of the macroscale BVP for a symmetrically notched dog-bone specimen. The bottom boundary is fixed in both the $x$ and $y$ directions, and a vertical displacement is applied to the top boundary.}
    \label{fig:13}
\end{figure}

\begin{figure}[h]
    \centering
    \includegraphics[keepaspectratio=true]{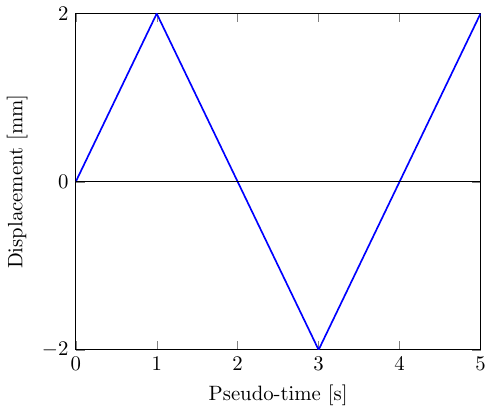}
    \caption{Cyclic displacement applied to the top boundary of the symmetrically notched dog-bone specimen plotted vs. (pseudo)-time.}
    \label{fig:14}
\end{figure}

To further explore the feasibility of utilizing the proposed homogenization approach for more complex and practical structural configurations, a symmetrically notched dog-bone specimen with the dimensions and BCs presented in Figure \ref{fig:13} is analyzed as the last numerical example. As a mixed isotropic and kinematic hardening rule is introduced in Section \ref{sec:4} for the plastic behavior, cyclic loading is applied to the specimen to assess if the current homogenization approach is capable of tracking the load path in the unloading and reverse loading regimes as well. To do so, the lower boundary of the dog-bone specimen is fixed in both the $x$ and $y$ directions, while a cyclic displacement in the $y$ direction is applied to the upper boundary. To better illustrate the sequence of events, the applied cyclic displacement to the top edge $u_y$ is plotted vs. pseudo-time in Figure \ref{fig:14}. The two-scale homogenization model is analyzed by using the sufficiently fine macroscopic FE discretization presented in Figure \ref{fig:13}.

\begin{figure}[h]
    \centering
    \includegraphics[keepaspectratio=true]{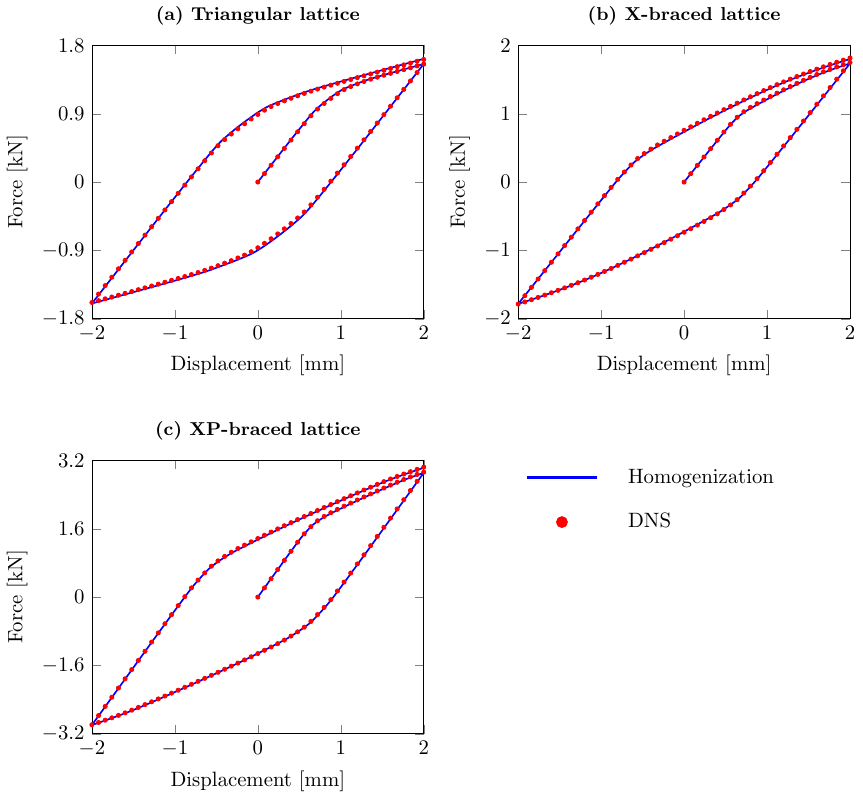}
    \caption{Force-displacement curves of the symmetrically notched dog-bone specimen for the \textbf{a)} triangular lattice, \textbf{b)} X-braced lattice and \textbf{c)} XP-braced lattice.}
    \label{fig:15}
\end{figure}

The plots depicting the sum of forces on the top boundary vs. the applied displacement are presented in Figure \ref{fig:15} for all the considered unit cell types. It can be seen that the homogenization results are in close agreement with those of DNS. Such an agreement is observed not only in the loading zone but also in the unloading and reverse loading regions. More specifically, it is observed that the adopted homogenization approach can quite accurately exhibit the kinematic hardening behavior to mimic the Bauschinger effect.

\begin{figure}[h]
    \centering
    \includegraphics[keepaspectratio=true, width = 0.75\linewidth]{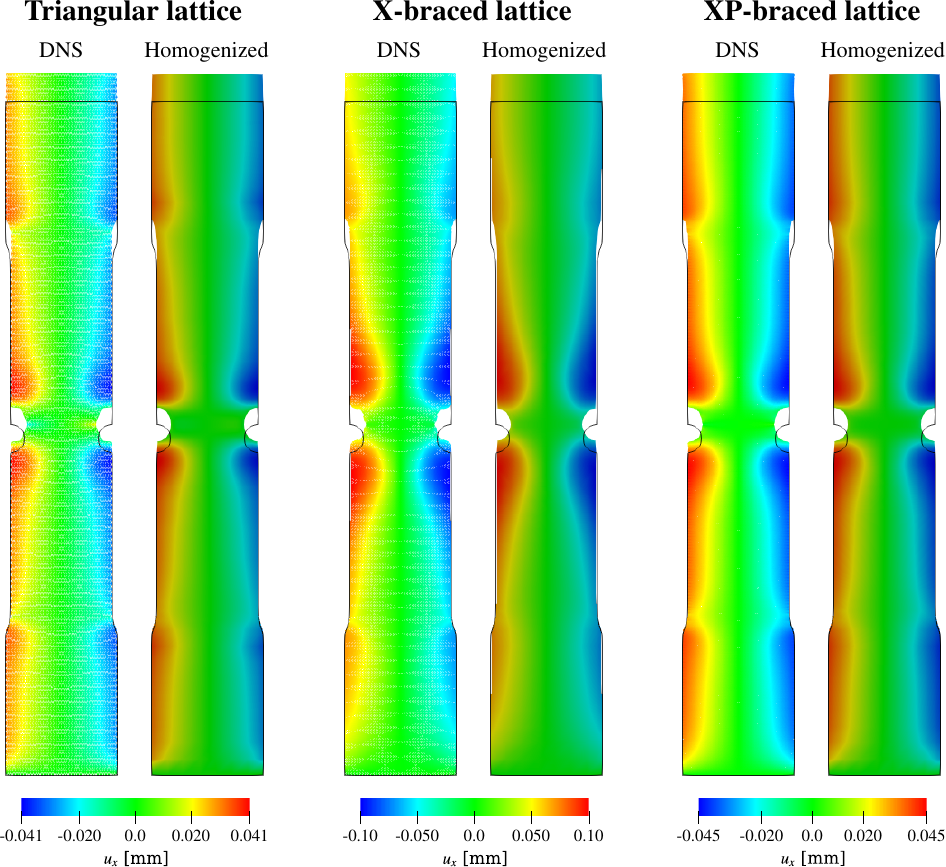}
    \caption{Displacement distribution in the $x$ direction for the symmetrically notched dog-bone specimen computed from homogenization and DNS for the triangular, X-braced and XP-braced lattices.}
    \label{fig:16}
\end{figure}

Finally, in pursuit of a more detailed comparison, the displacement distribution in the $x$ direction $u_x$ obtained from both DNS and homogenization is presented in Figure \ref{fig:16} for the three unit cell types. It is observed that for all the studied lattice topologies, the employed homogenization scheme yields a displacement distribution closely analogous to that of DNS throughout the specimen. Therefore, it is concluded that this homogenization approach is capable of providing sufficiently accurate results not only from the overall structural behavior point of view but also from a more local perspective.

\section{Conclusions}
\label{sec:7}

In the present work, a two-scale computational homogenization framework with on-the-fly transition of information was adopted to study the elastoplastic behavior of stretching-dominated truss-based metamaterials. Two BVPs were considered at the macroscale and the microscale to represent the overall homogenized structure and the underlying lattice structures, respectively. The macroscopic BVP was solved by the FE method with four-node bi-linear quadrilateral elements, while the microscale lattice structures were discretized by two-node linear displacement truss elements. The transition between the two scales was performed through the Hill-Mandel macrohomogenity condition. The macroscopic strain field was transferred to the microscale, where it was employed to compute the nodal displacements and, consequently, the axial strain of the lattice struts. An elastoplastic material behavior with nonlinear Voce isotropic hardening and linear kinematic hardening was considered for the microscale struts. Following the computation of the unit cell stiffness matrix by employing the return mapping algorithm for each lattice strut, the homogenized stress and tangent stiffness were transferred back to the macroscale.

The performance of the homogenization approach was investigated through a number of numerical examples considering three different unit cell topologies, including the triangular, X-braced and XP-braced lattices. The mesh convergence behavior as well as the fulfilment of the principle of scale separation were studied for a double-clamped beam. It was shown that by using a sufficiently large number of lattice structures, the homogenization framework provides force-displacement curves in close agreement with those obtained from full-scale simulations of the discrete structure. In the second numerical example, a square-shaped plate under tensile loading was considered, where it was shown that in addition to the force-displacement curves, the employed homogenization approach can also almost precisely track the displacement of the boundary transverse to the direction of loading. Moreover, the effective Poisson's ratios of the studied lattice topologies were computed. It was observed that the X-braced lattice exhibits the highest effective elastic Poisson’s ratio, while such a value is the lowest for the triangular lattice when loaded in the $x$ direction. The principle of scale separation was examined for the triangular lattice under $y$-direction loading by considering an equivalent elastoplastic Poisson's ratio as the basis for assessment. It was once again shown that as the number of considered unit cells increased, the equivalent elastoplastic Poisson's ratio obtained from homogenization was closer to that of DNS. Ultimately, a more complex structure, i.e., a dog-bone specimen with symmetric notches under cyclic loading, was considered as the last numerical study. The homogenization scheme was shown to be able to follow the same force-displacement path as DNS not only when the structure is monotonically loaded but also when it experiences unloading and reverse loading. In general, for all three numerical examples and all three lattice families, it was observed that the accuracy of the homogenization framework is slightly higher in the elastic regime, and as the structure undergoes more plastic deformation, such accuracy is slightly decreased. However, with a sufficient number of unit cells, the adopted homogenization approach provided precise enough solutions in both the elastic and elastoplastic zones.

Finally, future studies may proceed by extending the present model to finite deformations and rotations through corotational beam formulations \cite{crisfield1990consistent}, similar to the approach adopted for elastic \cite{glaesener2019continuum,glaesener2020continuum} and viscoelastic \cite{glaesener2021viscoelastic} truss-based metamaterials. Furthermore, enhancing the model by considering more complicated material behaviors, such as damage, may be of great interest, especially for metallic lattice structures, which experience a mixed elastic-plastic-damage behavior \cite{geng2019fracture,babamiri2020deformation,drucker2021experimental}. In the present work, the geometrical dimensions of the lattice struts were considered in such a way that buckling would not happen within the range of applied loadings. Accounting for such instabilities of lattice struts in combination with plasticity, similar to the approach employed in \cite{vigliotti2014non} for elastic lattices, would be insightful for the precise modeling of truss-based metamaterials. This requires the implementation of a robust solver capable of detecting bifurcation points and switching to the most critical branch to follow the lowest equilibrium path (e.g., see \cite{crisfield1981fast,de2012nonlinear,vigliotti2014non}). Topology optimization in a two-scale computational homogenization framework has already been considered in \cite{telgen2022topology} for elastic lattice structures. Topology and size optimization studies beyond the proportional limit within the presented two-scale framework would be interesting to perform. In the present work, the number of DoFs was significantly decreased at the microscale by utilizing truss elements instead of discretizing the lattice structures with continuum solid elements. The computational cost of the present two-scale approach can be further decreased by employing reduced integration elements with hourglass stabilization at the macroscale, such as the well-established elements developed in \cite{reese2003consistent,reese2005physically,barfusz2021single}. Another interesting advancement to further speed up computational time is the possibility of employing machine learning approaches. Among others, the recently developed physics-informed neural networks (PINNs) \cite{raissi2019physics,rezaei2022mixed,harandi2024mixed} have been shown in \cite{rezaei2024learning}, specifically for truss structures, to have a speed-up factor one order of magnitude faster than the implicit return mapping algorithm while maintaining accuracy. Replacing the conventional return mapping algorithm employed in the present work with such a machine learning approach can significantly improve the computational time by bypassing the repetitive iterations required to solve the nonlinear elastoplastic equations for each strut.
\begin{appendices}

\numberwithin{equation}{section}

\section{Topology matrices for the X-braced and XP-braced lattices}
\label{app:a}

Following the node numbering presented in Figure \ref{fig:2} and considering that the periodic vectors for the X-braced and XP-braced lattices are identical and written in terms of lattice side length $L_{uc}$ as 

\begin{equation}
\label{eq:a1}
    \bm{a}_{1} = L_{uc} \begin{pmatrix}
    1\\
    0
    \end{pmatrix},
    \quad
    \bm{a}_{2} = L_{uc} \begin{pmatrix}
    0\\
    1
    \end{pmatrix},
\end{equation}

the topology matrices for the X-braced and XP-braced lattices can be extracted as follows.

With respect to the X-braced lattice (Figure \ref{fig:2}b), the nodal positions can be written as

\begin{align}
\begin{split}
\label{eq:a2}
    & \bm{r}_{3} = \bm{r}_{2} + \bm{a}_{1},\\
    & \bm{r}_{4} = \bm{r}_{2} + \bm{a}_{2},\\
    & \bm{r}_{5} = \bm{r}_{2} + \bm{a}_{1} + \bm{a}_{2}.
\end{split}
\end{align}

Then, after the deformation, the nodal displacement are expressed as

\begin{align}
\begin{split}
\label{eq:a3}
    & \bm{d}_{3} = \bm{d}_{2} + \bar{\bm{\varepsilon}} \bm{a}_{1},\\
    & \bm{d}_{4} = \bm{d}_{2} + \bar{\bm{\varepsilon}} \bm{a}_{2},\\
    & \bm{d}_{5} = \bm{d}_{2} + \bar{\bm{\varepsilon}} \left(\bm{a}_{1} + \bm{a}_{2} \right),
\end{split}
\end{align}

which, based on the matrix representation of Eq. \ref{eq:13}, results in the following topology matrices:

\begin{equation}
\label{eq:a4}
    \bm{d} = \begin{pmatrix}
    \bm{d}_{1}\\
    \bm{d}_{2}\\
    \bm{d}_{3}\\
    \bm{d}_{4}\\
    \bm{d}_{5}
    \end{pmatrix}
\end{equation}

\begin{equation}
\label{eq:a5}
    \bm{d}_0 = \begin{pmatrix}
    \bm{d}_{1}\\
    \bm{d}_{2}
    \end{pmatrix}
\end{equation}

\begin{equation}
\label{eq:a6}
    \bm{B}_{0} = \begin{pmatrix}
    \bm{I} & \bm{0}\\
    \bm{0} & \bm{I}\\
    \bm{0} & \bm{I}\\
    \bm{0} & \bm{I}\\
    \bm{0} & \bm{I}
    \end{pmatrix}
\end{equation}

\begin{equation}
\label{eq:a7}
    \bm{B}_{e} = \begin{pmatrix}
    \bm{0}\\
    \bm{0}\\
    \bm{B}_{{e}_3}\\
    \bm{B}_{{e}_4}\\
    \bm{B}_{{e}_5}
    \end{pmatrix}
\end{equation}

\begin{equation}
\label{eq:a8}
    \bm{B}_{e_{3}} = \begin{pmatrix}
    a_{1x} & 0 & a_{1y} \\
    0 & a_{1y} & a_{1x}
    \end{pmatrix}
\end{equation}

\begin{equation}
\label{eq:a9}
    \bm{B}_{e_{4}} = \begin{pmatrix}
    a_{2x} & 0 & a_{2y} \\
    0 & a_{2y} & a_{2x}
    \end{pmatrix}
\end{equation}

\begin{equation}
\label{eq:a10}
    \bm{B}_{e_{5}} = \begin{pmatrix}
    a_{1x} + a_{2x} & 0 & a_{1y} + a_{2y} \\
    0 & a_{1y} + a_{2y} & a_{1x} + a_{2x}
    \end{pmatrix}
\end{equation}

Regarding the XP-braced lattice (Figure \ref{fig:2}c), the nodal positions will have the form

\begin{align}
\begin{split}
\label{eq:a11}
    & \bm{r}_{5} = \bm{r}_{2} + \bm{a}_{1},\\
    & \bm{r}_{6} = \bm{r}_{2} + \bm{a}_{2},\\
    & \bm{r}_{7} = \bm{r}_{2} + \bm{a}_{1} + \bm{a}_{2},\\
    & \bm{r}_{8} = \bm{r}_{3} + \bm{a}_{1},\\
    & \bm{r}_{9} = \bm{r}_{4} + \bm{a}_{2},\\
\end{split}
\end{align}

which yields the following expressions for the nodal displacement:

\begin{align}
\begin{split}
\label{eq:a12}
    & \bm{d}_{5} = \bm{d}_{2} + \bar{\bm{\varepsilon}} \bm{a}_{1},\\
    & \bm{d}_{6} = \bm{d}_{2} + \bar{\bm{\varepsilon}} \bm{a}_{2},\\
    & \bm{d}_{7} = \bm{d}_{2} + \bar{\bm{\varepsilon}} \left(\bm{a}_{1} + \bm{a}_{2} \right),\\
    & \bm{d}_{8} = \bm{d}_{3} + \bar{\bm{\varepsilon}} \bm{a}_{1},\\
    & \bm{d}_{9} = \bm{d}_{4} + \bar{\bm{\varepsilon}} \bm{a}_{2}.
\end{split}
\end{align}

Then, following the matrix representation of Eq. \ref{eq:13}, the topology matrices for the XP-braced lattice can be expressed as follows:

\begin{equation}
\label{eq:a13}
    \bm{d} = \begin{pmatrix}
    \bm{d}_{1}\\
    \bm{d}_{2}\\
    \bm{d}_{3}\\
    \bm{d}_{4}\\
    \bm{d}_{5}\\
    \bm{d}_{6}\\
    \bm{d}_{7}\\
    \bm{d}_{8}\\
    \bm{d}_{9}
    \end{pmatrix}
\end{equation}

\begin{equation}
\label{eq:a14}
    \bm{d}_0 = \begin{pmatrix}
    \bm{d}_{1}\\
    \bm{d}_{2}\\
    \bm{d}_{3}\\
    \bm{d}_{4}
    \end{pmatrix}
\end{equation}

\begin{equation}
\label{eq:a15}
    \bm{B}_{0} = \begin{pmatrix}
    \bm{I} & \bm{0} & \bm{0} & \bm{0}\\
    \bm{0} & \bm{I} & \bm{0} & \bm{0}\\
    \bm{0} & \bm{0} & \bm{I} & \bm{0}\\
    \bm{0} & \bm{0} & \bm{0} & \bm{I}\\
    \bm{0} & \bm{I} & \bm{0} & \bm{0}\\
    \bm{0} & \bm{I} & \bm{0} & \bm{0}\\
    \bm{0} & \bm{I} & \bm{0} & \bm{0}\\
    \bm{0} & \bm{0} & \bm{I} & \bm{0}\\
    \bm{0} & \bm{0} & \bm{0} & \bm{I}
    \end{pmatrix}
\end{equation}

\begin{equation}
\label{eq:a16}
    \bm{B}_{e} = \begin{pmatrix}
    \bm{0}\\
    \bm{0}\\
    \bm{0}\\
    \bm{0}\\
    \bm{B}_{{e}_5}\\
    \bm{B}_{{e}_6}\\
    \bm{B}_{{e}_7}\\
    \bm{B}_{{e}_8}\\
    \bm{B}_{{e}_9}
    \end{pmatrix}
\end{equation}

\begin{equation}
\label{eq:a17}
    \bm{B}_{e_{5}} = \bm{B}_{e_{8}} = \begin{pmatrix}
    a_{1x} & 0 & \frac{1}{2} a_{1y} \\
    0 & a_{1y} & \frac{1}{2} a_{1x}
    \end{pmatrix}
\end{equation}

\begin{equation}
\label{eq:a18}
    \bm{B}_{e_{6}} = \bm{B}_{e_{9}} = \begin{pmatrix}
    a_{2x} & 0 & \frac{1}{2} a_{2y} \\
    0 & a_{2y} & \frac{1}{2} a_{2x}
    \end{pmatrix}
\end{equation}

\begin{equation}
\label{eq:a19}
    \bm{B}_{e_{7}} = \begin{pmatrix}
    a_{1x} + a_{2x} & 0 & \frac{1}{2} \left( a_{1y} + a_{2y} \right) \\
    0 & a_{1y} + a_{2y} & \frac{1}{2} \left( a_{1x} + a_{2x} \right)
    \end{pmatrix}
\end{equation}

\end{appendices}
\section*{Acknowledgements}
\setlength\intextsep{1.25em}

\begin{wrapfigure}[4]{l}{6.25em}
    \includegraphics[keepaspectratio=true, height=4.1em]{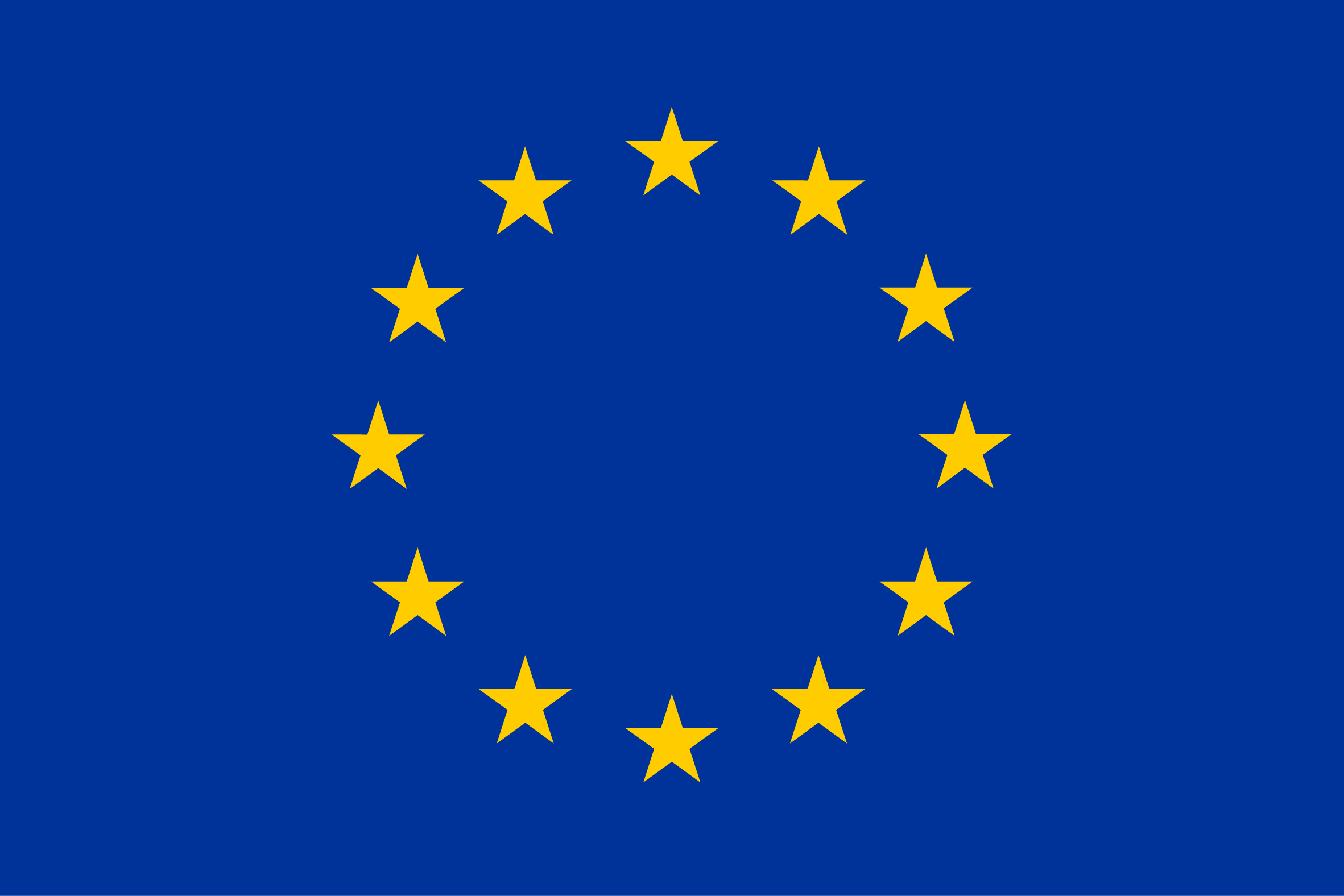}
\end{wrapfigure}
\, \\
\noindent The authors gratefully acknowledge the funding from the European Union’s Horizon 2020 research and innovation program under the Marie Skłodowska-Curie grant agreement No 956401 (XS-Meta).


\printbibliography

\end{document}